\documentclass[letterpaper,english,reprint, aps]{revtex4-1}
\usepackage[LGR,T1]{fontenc}
\usepackage[latin9]{inputenc}
\usepackage{array}
\usepackage{multirow}
\usepackage{amsmath}
\usepackage{amssymb}
\usepackage{graphicx}
\PassOptionsToPackage{version=3}{mhchem}
\usepackage{mhchem}

\usepackage{amsmath}
\let\oldAA\AA
\renewcommand{\AA}{\text{\normalfont\oldAA}}

\makeatletter

\ProvideTextCommand{\~}{LGR}[1]{\char126#1}

\usepackage{physics}

\makeatother

\usepackage{babel}
\begin{document}
\preprint{APS/123-QED}
\title{Confinement-Induced Isosymmetric Metal-Insulator Transition in \\
Ultrathin Epitaxial V$_{2}$O$_{3}$ Films}
\author{Simon Mellaerts\textsuperscript{1}, Claudio Bellani\textsuperscript{2},
Wei-Fan Hsu\textsuperscript{1}, Alberto Binetti\textsuperscript{1},
Koen Schouteden\textsuperscript{1}, Maria Recaman-Payo\textsuperscript{1},
Mariela Menghini\textsuperscript{1,3}, Juan Rubio Zuazo\textsuperscript{4,5},
Jes\'us L\'opez S\'anchez\textsuperscript{4,5,6}, Jin Won Seo\textsuperscript{2},
Michel Houssa\textsuperscript{1,7} and Jean-Pierre Locquet\textsuperscript{1}}
\affiliation{\textsuperscript{1}Department of Physics and Astronomy, KU Leuven,
Celestijnenlaan 200D, 3001 Leuven, Belgium, \textsuperscript{2}Department
of Materials Engineering, KU Leuven, Kasteelpark Arenberg 44, 3001
Leuven, Belgium, \textsuperscript{3}IMDEA Nanociencia, Calle Faraday
9, E29049 Madrid, Spain, \textsuperscript{4}BM25-SpLine, ESRF, France, \textsuperscript{5}Instituto de Ciencia de Materiales de Madrid (ICMM-CSIC), 28049 Madrid, Spain, \textsuperscript{6}Departamento de Electrocer\'amica, Instituto de Cer\'amica y Vidrio - Consejo Superior de Investigaciones Cient\'ificas (ICV-CSIC), Calle Kelsen 5, 28049 Madrid, Spain, \textsuperscript{7}Imec,
Kapeldreef 75, 3001 Leuven, Belgium.}
\email{simon.mellaerts@kuleuven.be}

\begin{abstract}
Dimensional confinement has shown to be an effective strategy to tune competing
degrees of freedom in complex oxides. Here, we achieved atomic layered growth of trigonal vanadium sesquioxide ($\ce{V}_{2}\ce{O}_{3}$) by means of oxygen-assisted molecular beam epitaxy. This led to a series of high-quality epitaxial ultrathin $\ce{V}_{2}\ce{O}_{3}$ films down to unit cell thickness, enabling the study of the intrinsic electron correlations upon confinement.
By electrical and optical measurements, we demonstrate a dimensional
confinement-induced metal-insulator transition in these
ultrathin films. We shed light on the Mott-Hubbard
nature of this transition, revealing an abrupt vanishing of the quasiparticle
weight as demonstrated by photoemission spectroscopy. Furthermore, we prove that dimensional
confinement acts as an effective out-of-plane stress. This highlights the structural component of correlated oxides in a confined architecture, while opening
an avenue to control both in-plane and out-of-plane lattice components
by epitaxial strain and confinement, respectively.
\end{abstract}
\maketitle

\section{\label{sec:Intro}Introduction}

The richness of transition metal oxides emerges from the strong interplay
of many degrees of freedom leading to competing ground states, whose
energy landscape can be deformed by temperature, pressure, and many
other external parameters. The room-temperature (RT) metal-insulator
transition (MIT) in vanadium sesquioxide ($\ce{V}_{2}\ce{O}_{3}$)
is a key example of an isosymmetric Mott transition which can be induced
by pressure, temperature, epitaxial strain and Cr doping \citep{McWhan1970PressureV2O3,McWhan,Rodolakis2010,Pia2017Collapse,Pia2021APLMaterial}.
For the latter approach, extended X-ray absorption fine-structure
spectroscopy (EXAFS) has shown that this RT MIT involves a local trigonal
distortion that results in a long-range strain driving an abrupt change
in the $c/a$ ratio \citep{McWhan,Grieger2014EffectCrdoping}.

On an electronic level, the octahedral crystal field (CF) of V $3d$
with a lower triple-degenerate $t_{2g}$ and upper two-fold degenerate
$e_{g}^{\sigma}$ orbital levels undergoes a symmetry lowering of
the former $t_{2g}$ level into an $a_{1g}$ singlet pointing along
the $c$ axis and a lower $e_{g}^{\pi}$ doublet in the basal plane
due to the trigonal distortion. This distortion leads to an enhanced
CF splitting between $a_{1g}$ and $e_{g}^{\pi}$ driving the electronic
MIT, as also shown by dynamical mean-field theory (DMFT) calculations
\citep{Poteryaev2007EnhancedCF}. 

With the advances in high-quality and layer-by-layer epitaxial growth
of ultrathin films and heterostructures, a vast playground has become available to tune these orbital degrees of freedom, which are extremely sensitive
to hybridization, CF and the local atomic environment \citep{Zoology2021,AnisotropyCorrelations2021}.
Ultimately, this opens the possibility towards the stabilization of a single atomic layer
of $\ce{V}_{2}\ce{O}_{3}$ with the promise of a high-temperature
ferromagnetic Chern insulator \citep{2DV2O3Simon}.

\

In this work, we propose dimensional confinement as an alternative
pressure term to tune the RT MIT in $\ce{V}_{2}\ce{O}_{3}$. High-quality
ultrathin $\ce{V}_{2}\ce{O}_{3}$ films were grown coherently on $\ce{Al}_{2}\ce{O}_{3}$
$(0001)$ substrates with thickness ranging from $6-18$ monolayers
(MLs) ($1.4-4.2$ nm) by the use of oxygen-assisted molecular beam
epitaxy (MBE). It is shown that the out-of-plane confinement in these
ultrathin films induces an intrinsic isosymmetric MIT at RT. By the
use of photoemission spectroscopy (PES), we prove the bandwidth-controlled
nature of this transition with an abrupt vanishing of the quasiparticle
(QP) weight at the Fermi level. Subsequently, Raman spectroscopy and
synchrotron X-ray diffraction show the stress-induced nature of this transition.

\section{\label{sec:Methods} methods}

All thin films were deposited on $(0001)$-$\ce{Al}_{2}\ce{O}_{3}$
by means of oxygen-assisted MBE in ultrahigh vacuum (UHV) conditions
with the growth chamber at base pressure of $10^{-10}$ mbar. MBE
growth was monitored in-situ by the use of reflection high-energy
electron diffraction (RHEED).

Chemical characterization was performed (without exposure to ambient
air) by X-ray photoelectron spectroscopy (XPS) in an UHV FlexMod SPECS
system equipped with a monochromatic $100\:$ W Al source ($E=1486.7$
eV), operating at a base pressure $10^{-10}$ mbar. The XPS spectra
were recorded with stepsize $0.05$ eV and pass energy $20$ eV. The
morphology of the ultrathin films was characterized by atomic force
microscopy (AFM) (ParkXE-100 AFM) under ambient conditions at RT. AFM
images were recorded in non-contact mode with the use of Si probes.
The structural properties of the samples were characterized by means
of X-ray reflection (XRR) and reciprocal space mapping (RSM) with
a Panalytical X'pert Pro diffractometer using a $\ce{Cu}$ anode with
$K_{\alpha1}$ radiation.

\

Temperature-dependent resistivity measurements in the Van der Pauw
configuration were performed in an Oxford Optistat CF2-V cryostat
with a Keithley 4200-SCS parameter analyzer. Simultaneous Fourier-transform
infrared (FTIR) spectroscopy measurements with a Bruker Vertex V80
were taken as a function of temperature, controlled by a thermocoupler.
Angle-resolved ultraviolet photoelectron spectroscopy (ARUPS) was
performed in the UHV FlexMod SPECS system at RT using a UVS300 high-intensity
vacuum ultraviolet light source optimized for He-II radiation ($E=40.8$
eV). The ARUPS spectra were recorded with stepsize of $0.01$ eV and
pass energy of $30$ eV. 
Raman spectra were acquired by a confocal Raman microscope Witec ALPHA model 300RA (Oxford instruments, Abingdon, UK) with a Nd:YAG green laser source of 532 nm in p-polarization. Samples were placed on a piezo-driven scan platform with a high positioning accuracy of 4 nm and 0.5 nm in lateral and vertical directions, respectively. Raman scans were carried out at RT using a $100\times$ objective with a numerical aperture of 0.95. The output laser power was $40$ mW to minimize overheating effects and sample damage. Spectra were collected in the spectral range 65-3850 cm$^{-1}$ by using a 600 g mm$^{-1}$ grating. Subsequently, raw data were processed by the Witec Project Plus software (version 2.08). 

\

Synchrotron radiation high-resolution X-Ray diffraction were performed at the SpLine CRG BM25 beamline at the ESRF The European Synchrotron (Grenoble, France). The wavelength of the X-ray beam was $0.496\,\AA$. Measurements were carried out in reflection geometry using a six-circle diffractometer in vertical configuration. XRD patterns were collected between the range of 5$^\circ$ - 45$^\circ$ (2Q) using a 2D photon-counting X-ray MAXIPIX detector \citep{synchrotronXRD}.
\section{\label{sec:Results}Results}

\subsection{\label{subsec:Structural}High-quality growth of epitaxial ultrathin
films}

\begin{figure*}
\includegraphics[scale=0.72]{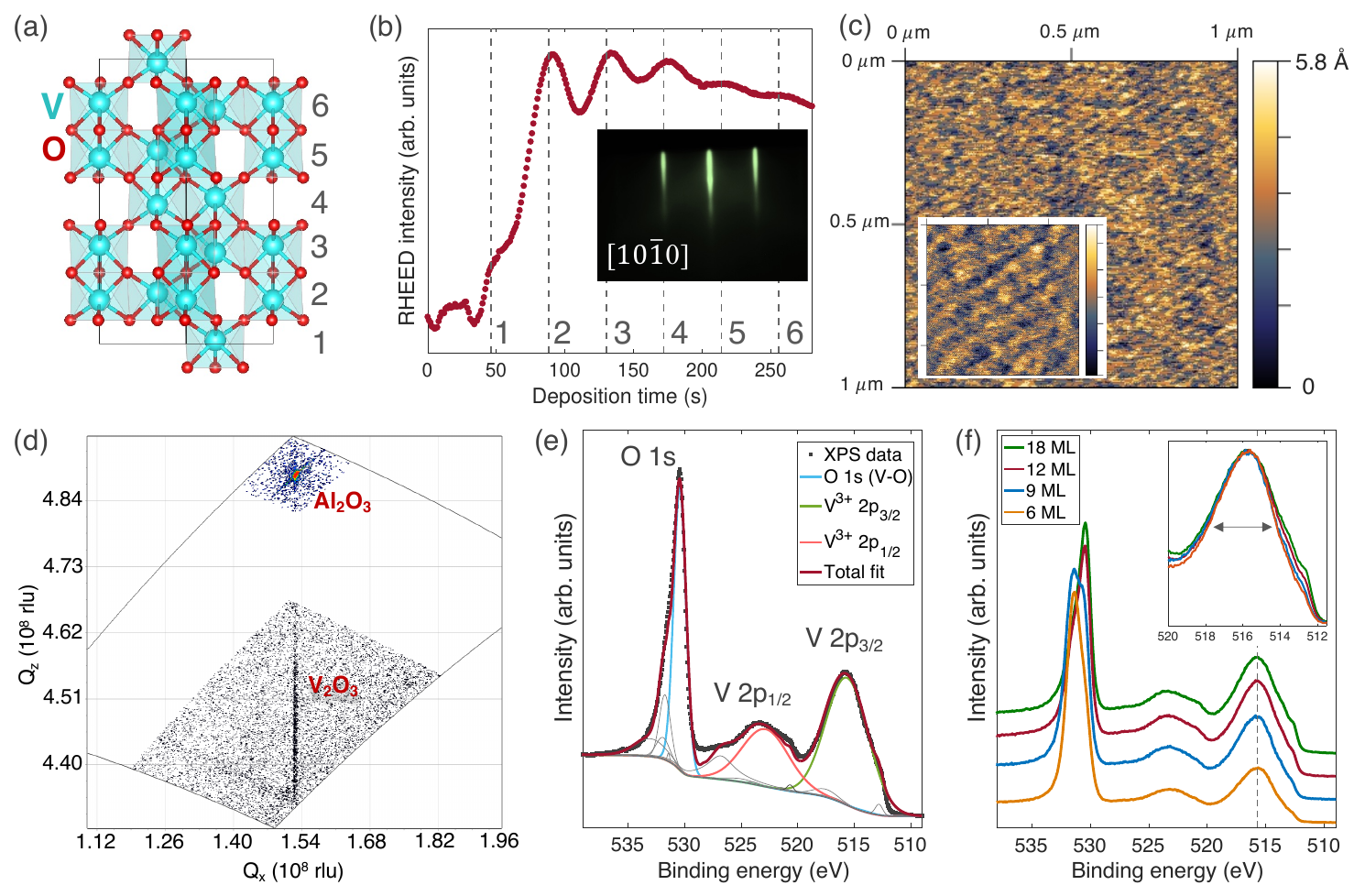}\caption{Structural and chemical properties of ultrathin films. (a) The conventional
unit cell of $\ce{V}_{2}\ce{O}_{3}$ consisting of six monolayers
with cyan and red colors corresponding to vanadium and oxygen atoms,
respectively. (b) RHEED intensity oscillations as a function of the
deposition time where each maximum corresponds to the completion of
one atomic layer and with the inset a RHEED pattern taken after the
growth. (c) AFM topography of the $18$ ML film with the inset an
AFM topography of dimension $500\times500$ nm$^2$ with a color scale
ranging from $0$ to $5.6\,\text{\AA}$. (d) RSM of the 18 ML film in the $(1\:0\,\bar{1}\,10)$
reflection showing a coherently strained film with identical in-plane
lattice constant to the $\ce{Al}_{2}\ce{O}_{3}$ substrate. (e) XPS
core level O $1s$ - V $2p$ spectrum of the $18$ ML with corresponding
fitting results for which the V$^{3+}$ components are highlighted
while the other stoichiometric components and satellites have been
gray colored. The details of the complete fitting result can be found
in Fig. S1 of the Supplemental Material. (f) A comparison of the core level O
$1s$ - V $2p$ spectra of the different ultrathin films. Note that
the O $1s$ level for the $6$ and $9$ ML thick films consists of
both a V-O as well as Al-O component. The inset shows the reducing
linewidth of the V $2p_{3/2}$ level upon thickness reduction.}
\label{Fig:Structural}
\end{figure*}

To study novel electronic phenomena in strongly correlated electron
systems where many energy scales compete in defining the ground state,
there is the need for the growth of high-quality films up to atomic
layer precision. In this respect, MBE has
shown to be one of the most successful compared to other deposition
techniques. By in-situ monitoring the RHEED pattern during the growth, we confirm the epitaxial
growth of the ultrathin films on atomically flat $(0001)$-$\ce{Al}_{2}\ce{O}_{3}$
substrates. In addition, intensity oscillations in the RHEED are observed
(see Figure \ref{Fig:Structural}b) with a periodicity corresponding
to the expected growth of one atomic layer, calibrated by the e-beam
flux and oxygen pressure. Hence, this corundum $\ce{V}_{2}\ce{O}_{3}$
material with conventional unit cell (UC) consisting of 6 ML (see
Figure \ref{Fig:Structural}a) can be grown in a ML fashion, with
the ability to control the thickness down to atomic layer precision. This ML growth is also evident from AFM images shown in Figure \ref{Fig:Structural}c
with estimated step terraces of width $32$
nm.

\ 

The lattice mismatch of the epitaxial $\ce{V}_{2}\ce{O}_{3}$ films
with the $\ce{Al}_{2}\ce{O}_{3}$ substrates leads to an in-plane compressive
strain of $4.2\:\%$. By RSM in $(1\:0\,\bar{1}\,10)$
reflection, it is shown that the ultrathin films are fully strained
in a coherent manner without any strain relaxation. This is due to
the fact that all films are below the critical thickness. Hence, in
the comparison between the different ultrathin films an identical
in-plane lattice constant can be assumed.

\ 

Vanadium oxides are known to have a rich phase diagram with
many different oxidation states to be stabilized \citep{Wriedt1989,Bahlawane2014}. 
Therefore, XPS was performed immediately after growth without exposure
to ambient air to prevent further oxidation of the films. The core
level O $1s$ - V $2p$ spectrum is shown in Figure \ref{Fig:Structural}e
with corresponding curve fitting, confirming the $\ce{V}^{3+}$ stoichiometry
(see Fig. S1 of the Supplemental Material for detailed analysis). Alternatively the
binding energy difference between O $1s$ and V $2p_{3/2}$ equals
$14.5$ eV, which also confirms this stoichiometry \citep{SILVERSMIT2004167}.
Also note that the broadened linewidth of the V $2p_{3/2}$ core state
can be ascribed to the presence of the electron correlations in the
$3d$, rather than a mixed oxidation state \citep{Sawatzky1979}.
Moreover, a comparison of core level XPS spectra of the different
ultrathin films proves that stoichiometry has been preserved for all
thicknesses (see Figure \ref{Fig:Structural}f). This excludes any
changes as a function of the thickness to be ascribed to a change
in the oxygen stoichiometry of the films. Nonetheless, a reducing linewidth
of the V $2p_{3/2}$ core level is observed upon thickness reduction
(see inset of Fig. \ref{Fig:Structural}f), suggesting the diminishment
of the electron correlations, which is indeed the case as will be
shown in the subsequent sections.

In contrast to earlier reports on ultrathin $\ce{V}_{2}\ce{O}_{3}$
films \citep{Luo2004,POLEWCZYK2023155462}, we prove an atomic layer
control in the growth of epitaxial coherent and stochiometric $\ce{V}_{2}\ce{O}_{3}$
in the ultrathin limit. Hence, this permits the study of the intrinsic
properties of this archetypical Mott material in the few MLs limit. 

\subsection{\label{subsec:Electronics}Electronics upon thickness reduction}

\begin{figure}[h]
\includegraphics[scale=0.51]{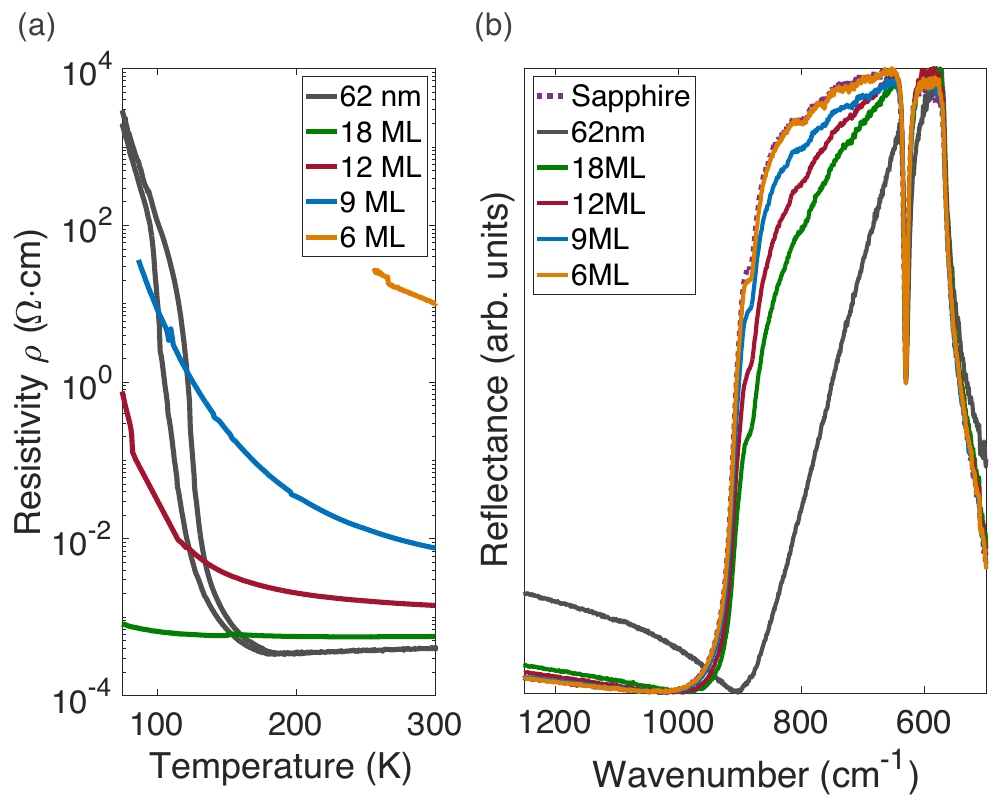}\caption{Electronics upon thickness reduction. (a) Temperature-dependent resistivity.
(b) Reflectance spectrum near the reststrahlen band of sapphire.}
\label{Fig:Electrical}
\end{figure}

The intrinsic properties of these ultrathin $\ce{V}_{2}\ce{O}_{3}$
films at RT were evaluated by the use of transport measurements combined
with infrared (IR) spectroscopy. Figure \ref{Fig:Electrical} shows
the temperature-dependent electrical resistivity $\rho(T)$ of the
ultrathin films. First of all, the coherently strained ultrathin films
do not show any LT MIT, which has been ascribed to the presence of
a large compressive strain and is therefore not related to thickness reduction \citep{Dillemans2014}.
The remainder of the manuscript focuses on the room-temperature properties
of these ultrathin films.

\begin{figure*}
\includegraphics[scale=0.62]{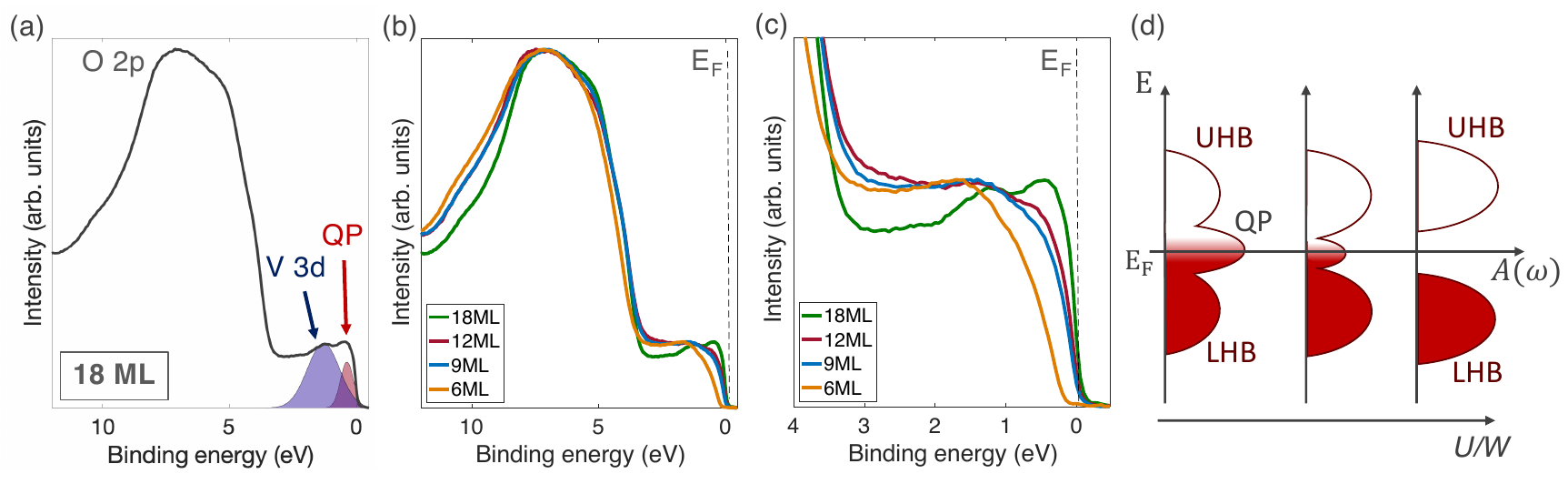}\caption{Photoemission spectroscopy. (a) Decomposition of the valence band
into $V\:3d$ band and QP weight for a film with $18$ ML thickness.
(b) A comparison of normalized PES spectra of the valence band where
(c) the expanded energy scale spectra shows a vanishing QP weight
upon thickness reduction. (d) A schematic view of the behavior of
the low-energy band structure upon increasing $U/W$ with 3 spectral components: the lower and upper Hubbard bands (LHB/UHB) and the quasiparticle (QP) weight.}
\label{figure:UPS}
\end{figure*}

As a reference, to prove the high-quality of the grown thin films,
$\rho(T)$ of a relaxed thin film of $62$ nm ($\sim44$ UC) is
included in Figure \ref{Fig:Electrical}a. This shows a sharp discontinuous
metal-insulator transition at $\sim180$ K while the room-temperature
resistivity (RTR) equal to $3.94\cdot10^{-4}$ $\Omega\cdot$cm agrees
very well with bulk values \citep{McWhan1970PressureV2O3,McWhan}.
Upon thickness reduction, the RTR value starts to increase, with an
abrupt increase below $9$ ML. The RTR of $6$ ML is nearly five times
larger than the $2.3$ $\Omega\cdot$cm RTR value observed in paramagnetic
insulating state stabilized for $4\%$ Cr-doped $\ce{V}_{2}\ce{O}_{3}$
single crystals \citep{McWhan}, and epitaxially strained thin films
\citep{Pia2021APLMaterial}. Also notice that the $18$ ML clearly
shows a metallic slope ($d\rho/dT>0$) at RT, which becomes semiconducting
below with an increasing slope upon confinement indicating an increased
electronic bandgap.

\

To support these transport measurements, the optical reflectance in
near-normal angle incidence is measured using FTIR spectroscopy. The
optical reflectance, shown in Figure \ref{Fig:Electrical}b, has been
studied in the mid-infrared range ($480-1250$ cm$^{-1}$), close to
the reststrahlen band of sapphire, that is the infrared band where
the reflectance is near-unity and the IR-active phonon resonances
appear. Any metallic thin film leads to a damping of this reststrahlen
effect, or thus absorbance by the film. It can be seen that this effect
reduces upon reducing thickness. Therefore, it can be concluded that
the metallicity dissapears below 9 ML thick ultrathin films, in agreement
with the transport measurements. A similar spectral change near the
reststrahlen band of sapphire can be observed upon LT MIT of the $60$
nm thin $\ce{V}_{2}\ce{O}_{3}$ film (See Fig. S3 in Supplemental
Material).

\subsection{\label{subsec:BWcontrolled}Bandwidth-controlled transition}

$\ce{V}_{2}\ce{O}_{3}$ is often considered as a key example of a
Mott-Hubbard system, that is a purely electronic MIT involving no
other degrees of freedom \citep{Imada1998,Held2001,Hansmann2013}. 

To evaluate the nature of the induced MIT upon thickness reduction,
PES was used to probe the changes in the valence band. These spectra
consist of a spectral weight derived from the O $2p$ states in the
energy range $4-10$ eV, while close to the Fermi level ($E_{F}$),
the valence band consists of a broad $V\:3d$ weight and a sharp coherent
QP weight, as shown in Figure \ref{figure:UPS}a. It should be pointed
out that the observation of the coherent QP weight required a high-quality
pristine surface, otherwise only a faint shoulder could be identified
(see Fig. S2 of the Supplemental Material), which indicates the importance of the
surface termination and texture. Moreover, the observation of such a large
QP weight in a $18$ ML ($\sim4.2$ nm) film is in excellent agreement
with the critical thickness estimate of $4$ nm required to see the QP weight,
as determined through angle-resolved PES experiments on single crystal
$\ce{V}_{2}\ce{O}_{3}$ \citep{Borghi2009QP,Rodolakis2009QP}.

\ 

Upon thickness reduction, the QP spectral weight starts to vanish
at $E_{F}$ while simulatenously the V $3d$ band shifts away from
$E_{F}$ opening an energy gap. This also leads to an energy shift
of the oxygen (O $2p$) band. The observed vanishing QP weight with
a spectral weight transfer to the V $3d$ band agrees with the bandwidth-controlled
Mott transition behavior derived by DMFT, schematically illustrated
in Figure \ref{figure:UPS}d \citep{Georges1996}.

This vanishing QP weight upon confinement can be understood by an
orbital-resolved analysis. It was shown that the coherent QP band
in the PM phase has a dominant $a_{1g}$ orbital character \citep{Rodolakis2009QP,Thees2021ARPES},
which transfers its spectral weight to the lower-lying $e_{g}^{\pi}$-derived
bands upon transition to the PI state \citep{Park2000occupation,Vecchio2015Optics,Thees2021ARPES}.
Now, reducing the thickness causes the electron localization in the
out-of-plane orbitals, including mainly $a_{1g}$. In a single-orbital
Hubbard model, this corresponds to a reducing effective bandwidth
$W$ that triggers a MIT at sufficiently large $U/W$ values with
$U$ the on-site Coulomb interactions. This scenario has been observed
in many other ultrathin oxide films, often referred to as a dimensional-crossover
\citep{Yoshimatsu2010SrVO3,Gu2013CaVO3,Sakai2013,Gu2014}.

\subsection{\label{subsec:Stress}Confinement as an effective stress component}

This confinement-induced (dimensional-crossover) MIT has also been
observed in vrious strongly correlated materials \citep{Yoshimatsu2010SrVO3,Sakai2013,Gu2014,Shiga2020}.
However, many of these studies have only considered the electronic
nature of this transition, because a structural characterization remains
challenging for these few nanometer thick films. In this respect, we have
used confocal Raman microscopy and synchrotron X-ray diffraction to
study the local interatomic lengths and lattice parameters, respectively.

\

\begin{figure}[h]
\includegraphics[scale=0.47]{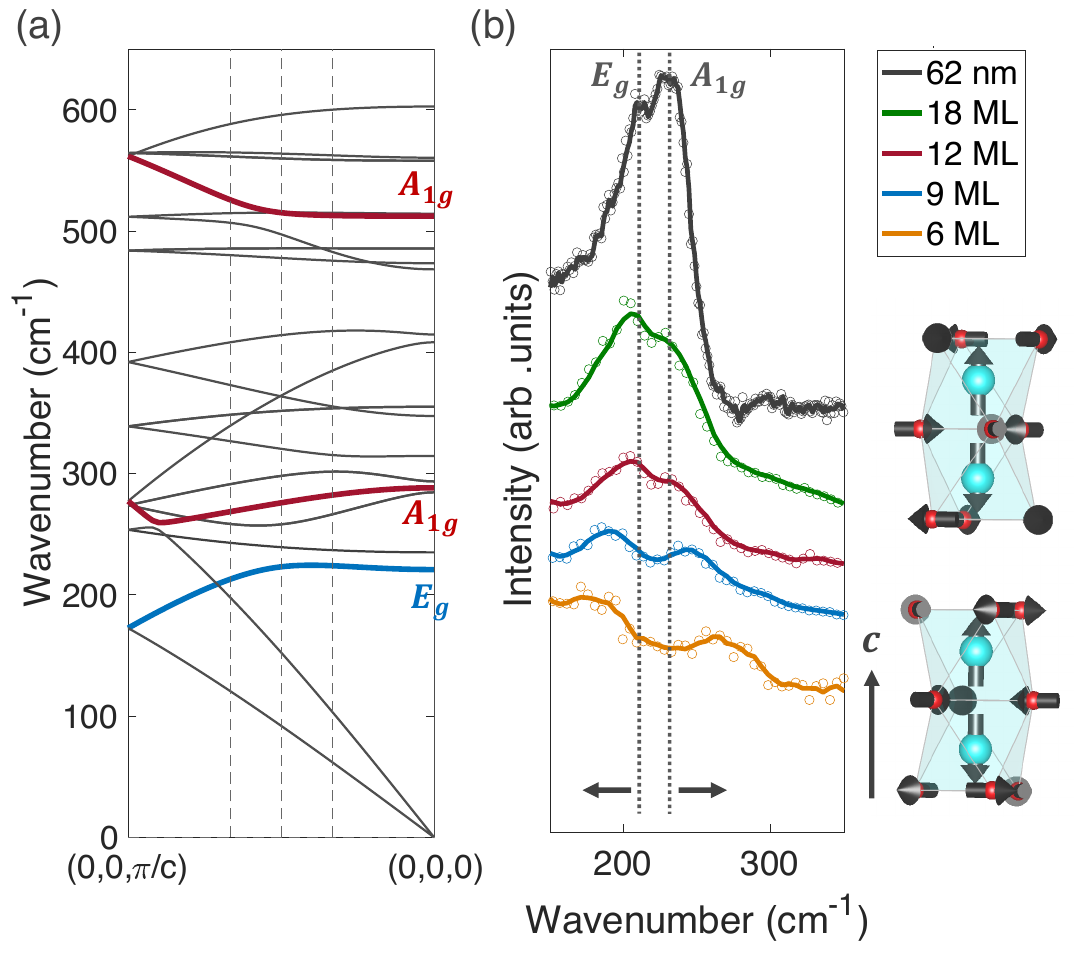}\caption{Phonon analysis as a function of thickness. (a) DFT-calculated phonon
branch along the $\langle001\rangle$ direction in the Brillouin zone
where $(0,0,0)$ corresponds to the zone-center $\Gamma$ and $(0,0,\pi/c)$
to the high-symmetry edge point $Z$. The phonon modes $A_{1g}$ and
$E_{g}$ under investigation are indicated in red and blue, respectively. The dashed lines represent $(0,0,\pi/d)$ with $d$ corresponding to 18, 12, and 9 ML from right to left.
(b) The Raman spectrum in the low-frequency region as a function of
thickness with on the right a visualization of the $A_{1g}$ mode
where the vectors correspond to the DFT-calculated eigenvectors of
the low-frequency vibration. For visual purposes the eigenvectors
have been scaled.}
\label{Fig:Raman}
\end{figure}

To monitor local atomic changes in the crystal structure, the Raman
active modes have been studied. For $\ce{V}_{2}\ce{O}_{3}$, it is
well-known that the $A_{1g}$ Raman mode, illustrated in Figure \ref{Fig:Raman}b,
can be used to probe the short V-V bond along $c$ axis, as well as
the trigonal distortion driving the RT MIT \citep{Kuroda1977,Poteryaev2007EnhancedCF,2023RamanV2O3}.
In particular, a frequency stiffening of the $A_{1g}$ mode observed
upon PM to PI MIT is ascribed to an elongated V-V bond where an interelectronic
Coulombic force is responsible for the stiffening of the atomic vibration
\citep{Yang1985,2023RamanV2O3}. Taking into account the observed
RT MIT in the ultrathin films, the question arises whether the trigonal
distortion and the corresponding V-V bond have also been altered upon
confinement.

Accordingly, the $A_{1g}$ Raman mode is studied as a function of
thickness in the ultrathin films, as shown in Figure \ref{Fig:Raman}.
To study changes in the phonon dynamics by confinement, the standing
wave approximation is proven to be a fruitful approach for many oxides
\citep{Senet1195,Premper2020,2020Schober}. For the trigonal $R\bar{3}c$
symmetry, the Raman active modes are non-polar and can be considered
to have a quantized wavevector $q$ along the out-of-plane Brillouin
zone branch ($\Gamma-Z$) due to confinement. More specifically, the
wavevector $q$ equals $(0,0,\pi/d)$ with $d$ the number of UCs (6
ML), with the bulk ($d=\infty$) corresponding to the zone-center
$\Gamma=(0,0,0)$ and the single UC (6ML) to the zone edge $Z=(0,0,\pi/c)$
with $c$ the out-of-plane lattice constant of the conventional UC.
This quantization of the wavevector $q$ also leads to a changing
eigenenergy, which can be obtained from the dispersion along the $\Gamma-Z$
branch. The latter was obtained by earlier density functional theory
(DFT) calculations \citep{2023RamanV2O3}, and is shown in Fig. \ref{Fig:Raman}a.

From the calculated dispersion a softening of the in-plane $E_{g}$
mode can be inferred upon confinement, while the low-frequency $A_{1g}$
shows an initial softening with an abrupt stiffening close to the
zone edge $Z$. The softening of the $E_{g}$ mode is confirmed by
Raman spectroscopy (see Fig. \ref{Fig:Raman}b), albeit that the DFT
calculations overestimates the frequency shift. This can be largely
explained by the approximation of a sharp boundary condition which
prohibits the propagation of the phonon mode across the interface
of the film with the substrate. In practice the similarity
in crystal structure and phonon spectrum can lead to a broadening
of this boundary condition, reducing the effective confinement. On
the other hand, a strong frequency hardening is observed for the $A_{1g}$
mode in the Raman spectrum which is inconsistent with the calculated
dispersion. This implies that the DFT calculations fail to describe
the V-V dimer behavior upon confinement, similar to the PM-PI MIT
in bulk $\ce{V}_{2}\ce{O}_{3}$ which can only be well-described by
DMFT \citep{Poteryaev2007EnhancedCF}. Moreover, the frequency hardening
of the $A_{1g}$ mode upon thickness reduction is very similar to
what has been observed in the PM-PI transition, which has been ascribed
to a trigonal distortion driving the MIT \citep{2023RamanV2O3}. 

\ 

At this point, the Raman spectroscopy results exclude any change in symmetry,
that means, the RT MIT is \textit{isosymmetric}. However, this does
not exclude an isostructural MIT as often misinterpreted. To verify
whether there is a structural change, synchrotron X-ray diffraction
was performed on 9 ML and 18 ML ultrathin films to determine the out-of-plane
$c$ lattice constant. A decrease of the $c$ lattice parameter is
observed upon thickness reduction from 18 ML to 9 ML, i.e. there is
a change in $c/a$ ratio, quantifying the trigonal distortion (See
Fig. S5 in Supplemental Material). The relative change of the $c/a$
between 18 ML and 9 ML equals $2.59\:\%$, which is larger
than the relative change of $1.32\:\%$ for the bulk PM-PI transition.
These results confirm the structural nature of this confinement-induced
MIT, where a trigonal distortion is triggered upon thickness reduction.
This out-of-plane strain component suggests that dimensional confinement
acts as a an alternative stress term to alter the trigonal distortion
triggering the MIT in $\ce{V}_{2}\ce{O}_{3}$. 

\section{\label{sec:Discussion}Discussion}

Although size confinement has been considered as a chemical pressure
in many nanostructures \citep{Lin2022ChemicalPreview}, such as nanowires
\citep{Sun2020NP-NanowiresPTO}, it remains unnoticed how confinement
in thin films and superlattices can lead to an effective out-of-plane
pressure. Therefore, we aim to provide an intuitive picture to perceive
dimensional confinement as an effective (chemical) pressure. Firstly,
by defining pressure in the material as $dP=-KdV/V$ with $K$ the
bulk modulus, it is evident that an induced pressure exists when there
is a change in volume $V$. In particular, considering $K>0$, a positive
pressure $P$ corresponds to $dV<0$, while $dV>0$ is associated
with a negative induced pressure. Reversely, changing the volume in
a controlled manner can lead to the inducement of a positive or negative
pressure/stress. Now, (chemical) pressure and chemical bonds are intimately
related, to the extent that (chemical) pressure has been used as
an effective way to identify and quantify the chemical bonds driving
the displacive phase transitions \citep{PhononsChemistryPV,2018ChemicalMaps}.
Therefore, we propose two views on how a change in pressure ($dP$)
or volume ($dV$) alters the electrons and their hybridization state.
(1) A positive pressure tends to narrow the electron density distribution
leading to an orbital expansion to decrease the energy, while negative
pressure is accompanied with interatomic delocalization in the regions
where covalent bonding occurs \citep{Osman2018CPmaps}. (2) Alternatively,
by approximating the kinetic (KE) and potential (PE) energies of electrons
by the volume dependences $V^{-2/3}$ and $V^{-1/3}$ respectively,
it becomes immediately evident that a change of volume, or thus a
pressure term, can be directly linked to a change in the hybridization
(PE/KE) of the electrons \citep{BTO2022Simon}.

In this way, pressure can be perceived as an effective way to alter
the electronic charge distribution and their corresponding bonds.
Reversely, altering the electronic charge distribution leads to a
(chemical) pressure within the crystal. Therefore, dimensional confinement
reducing the KE of the electrons occupying the out-of-plane orbitals
leads to an increased hybridization (PE/KE$\sim V^{1/3}$), and thus
also results in an effective negative pressure term. 

\ 

Accordingly, the structural changes (out-of-plane strain) discussed
in Section \ref{subsec:Stress} can be explained by an out-of-plane
stress that is strongly coupled to the enhanced hybridization of the
$a_{1g}$ orbital, which triggers the MIT. Similarly, many other physical
phenomena in ultrathin films and heterostructures can be explained as for example, the enhanced tetragonality in various perovskites, such
as $\ce{PbTiO}_{3}$, $\ce{BaTiO}_{3}$, and $\ce{SrTiO}_{3}$ \citep{2022UltrathinSTO}.
This involves the electron localization in the out-of-plane $\pi$-bonding
orbitals which favors the hybridization with the O $2p$ orbitals
strengthening the tetragonal distortion \citep{Thomann1987,Cohen1992}.
Similarly, it can be expected that the supression/enhancements of
octahedral rotations in perovskite-based heterostructures can be explained
by the same hybridization preference upon confinement \citep{CovalencyOctaTilts}.

Consequently, by combining epitaxial substrate-induced strain with
thickness control of the ultrathin films, a three-dimensional strain
can be imposed albeit that the out-of-plane component is uni-directional
whose sign depends on the electronic configuration.

\section{\label{sec:Conclusion} Conclusions}
The achievement of the atomic layered growth of this archetypical Mott material $\ce{V}_{2}\ce{O}_{3}$ has enabled the observation of an intrinsic MIT at RT induced by dimensional confinement. It has been identified as a clear illustration of a bandwidth-controlled Mott-Hubbard transition
with a vanishing QP below $18$ ML. 
Moreover, our findings highlight an enhanced trigonal distortion under confinement, emphasizing the structural
component of this MIT. The observed distortion suggests the possibility to model confinement along a specific direction as a negative stress component. Thereby illustrating the potential to leverage thickness control in ultrathin films or heterostructures to induce an effective out-of-plane stress.

\section*{Acknowledgements}
Part of this work was financially supported
by the KU Leuven Research Funds, Project No. C14/21/083, iBOF/21/084,
No. KAC24/18/056 and No. C14/17/080, as well as the FWO AKUL/13/19 and AKUL/19/023, and the Research Funds
of the INTERREG-E-TEST Project (EMR113) and INTERREG-VL-NL-ETPATHFINDER
Project (0559). Part of the computational resources and services used
in this work were provided by the VSC (Flemish Supercomputer Center)
funded by the Research Foundation Flanders (FWO) and the Flemish government. We acknowledge the use of synchrotron radiation facility at BM25-SpLine at the European Synchrotron and we thanks the Consejo Superior de Investigaciones Cientificas and the Ministerio de Ciencia e Innovacion for financial support under the proposal MA-5601 and projects PIE 2010 6 0E 013 and PIE 2021 60 E 030. J. S. L. acknowledges financial support from the Ministerio de Asuntos Econ\'omicos y Transformaci\'on Digital (MINECO) through the project PID2020-114192RB-C41. M. M. acknowledges support from Severo Ochoa Programme for Centres of Excellence in R\&D (MINCINN, Grant CEX2020-001039-S). 

\section*{Author contributions}
S.M., J.P., J.W.S. and M.H. conceived the conceptual idea and devised the research plan. S.M. executed the synthesis, while in-house diffraction, transport and optical experiments were carried out by S.M., C.B. and W.F.H. XPS, AFM and ARUPS were performed by K.S. Raman spectroscopy was conducted by J.S.L. Synchrotron X-ray diffraction was executed by S.M., W.F.H. and M.M. under guidance of J.S.L. and J.R.Z. All analysis and preparation of the manuscript were done by S.M.  All co-authors reviewed and revised the manuscript.

\section*{Competing interests}
The authors declare no competing interests.


\begin{thebibliography}{48}%
\makeatletter
\providecommand \@ifxundefined [1]{%
 \@ifx{#1\undefined}
}%
\providecommand \@ifnum [1]{%
 \ifnum #1\expandafter \@firstoftwo
 \else \expandafter \@secondoftwo
 \fi
}%
\providecommand \@ifx [1]{%
 \ifx #1\expandafter \@firstoftwo
 \else \expandafter \@secondoftwo
 \fi
}%
\providecommand \natexlab [1]{#1}%
\providecommand \enquote  [1]{``#1''}%
\providecommand \bibnamefont  [1]{#1}%
\providecommand \bibfnamefont [1]{#1}%
\providecommand \citenamefont [1]{#1}%
\providecommand \href@noop [0]{\@secondoftwo}%
\providecommand \href [0]{\begingroup \@sanitize@url \@href}%
\providecommand \@href[1]{\@@startlink{#1}\@@href}%
\providecommand \@@href[1]{\endgroup#1\@@endlink}%
\providecommand \@sanitize@url [0]{\catcode `\\12\catcode `\$12\catcode
  `\&12\catcode `\#12\catcode `\^12\catcode `\_12\catcode `\%12\relax}%
\providecommand \@@startlink[1]{}%
\providecommand \@@endlink[0]{}%
\providecommand \url  [0]{\begingroup\@sanitize@url \@url }%
\providecommand \@url [1]{\endgroup\@href {#1}{\urlprefix }}%
\providecommand \urlprefix  [0]{URL }%
\providecommand \Eprint [0]{\href }%
\providecommand \doibase [0]{http://dx.doi.org/}%
\providecommand \selectlanguage [0]{\@gobble}%
\providecommand \bibinfo  [0]{\@secondoftwo}%
\providecommand \bibfield  [0]{\@secondoftwo}%
\providecommand \translation [1]{[#1]}%
\providecommand \BibitemOpen [0]{}%
\providecommand \bibitemStop [0]{}%
\providecommand \bibitemNoStop [0]{.\EOS\space}%
\providecommand \EOS [0]{\spacefactor3000\relax}%
\providecommand \BibitemShut  [1]{\csname bibitem#1\endcsname}%
\let\auto@bib@innerbib\@empty
\bibitem [{\citenamefont {McWhan}\ \emph {et~al.}(1969)\citenamefont {McWhan},
  \citenamefont {Rice},\ and\ \citenamefont
  {Remeika}}]{McWhan1970PressureV2O3}%
  \BibitemOpen
  \bibfield  {author} {\bibinfo {author} {\bibfnamefont {D.~B.}\ \bibnamefont
  {McWhan}}, \bibinfo {author} {\bibfnamefont {T.~M.}\ \bibnamefont {Rice}}, \
  and\ \bibinfo {author} {\bibfnamefont {J.~P.}\ \bibnamefont {Remeika}},\
  }\href {\doibase 10.1103/PhysRevLett.23.1384} {\bibfield  {journal} {\bibinfo
   {journal} {Phys. Rev. Lett.}\ }\textbf {\bibinfo {volume} {23}},\ \bibinfo
  {pages} {1384} (\bibinfo {year} {1969})}\BibitemShut {NoStop}%
\bibitem [{\citenamefont {McWhan}\ and\ \citenamefont
  {Remeika}(1970)}]{McWhan}%
  \BibitemOpen
  \bibfield  {author} {\bibinfo {author} {\bibfnamefont {D.~B.}\ \bibnamefont
  {McWhan}}\ and\ \bibinfo {author} {\bibfnamefont {J.~P.}\ \bibnamefont
  {Remeika}},\ }\href {\doibase 10.1103/PhysRevB.2.3734} {\bibfield  {journal}
  {\bibinfo  {journal} {Phys. Rev. B}\ }\textbf {\bibinfo {volume} {2}},\
  \bibinfo {pages} {3734} (\bibinfo {year} {1970})}\BibitemShut {NoStop}%
\bibitem [{\citenamefont {Rodolakis}\ \emph {et~al.}(2010)\citenamefont
  {Rodolakis}, \citenamefont {Hansmann}, \citenamefont {Rueff}, \citenamefont
  {Toschi}, \citenamefont {Haverkort}, \citenamefont {Sangiovanni},
  \citenamefont {Tanaka}, \citenamefont {Saha-Dasgupta}, \citenamefont
  {Andersen}, \citenamefont {Held}, \citenamefont {Sikora}, \citenamefont
  {Alliot}, \citenamefont {Iti\'e}, \citenamefont {Baudelet}, \citenamefont
  {Wzietek}, \citenamefont {Metcalf},\ and\ \citenamefont
  {Marsi}}]{Rodolakis2010}%
  \BibitemOpen
  \bibfield  {author} {\bibinfo {author} {\bibfnamefont {F.}~\bibnamefont
  {Rodolakis}}, \bibinfo {author} {\bibfnamefont {P.}~\bibnamefont {Hansmann}},
  \bibinfo {author} {\bibfnamefont {J.-P.}\ \bibnamefont {Rueff}}, \bibinfo
  {author} {\bibfnamefont {A.}~\bibnamefont {Toschi}}, \bibinfo {author}
  {\bibfnamefont {M.~W.}\ \bibnamefont {Haverkort}}, \bibinfo {author}
  {\bibfnamefont {G.}~\bibnamefont {Sangiovanni}}, \bibinfo {author}
  {\bibfnamefont {A.}~\bibnamefont {Tanaka}}, \bibinfo {author} {\bibfnamefont
  {T.}~\bibnamefont {Saha-Dasgupta}}, \bibinfo {author} {\bibfnamefont {O.~K.}\
  \bibnamefont {Andersen}}, \bibinfo {author} {\bibfnamefont {K.}~\bibnamefont
  {Held}}, \bibinfo {author} {\bibfnamefont {M.}~\bibnamefont {Sikora}},
  \bibinfo {author} {\bibfnamefont {I.}~\bibnamefont {Alliot}}, \bibinfo
  {author} {\bibfnamefont {J.-P.}\ \bibnamefont {Iti\'e}}, \bibinfo {author}
  {\bibfnamefont {F.}~\bibnamefont {Baudelet}}, \bibinfo {author}
  {\bibfnamefont {P.}~\bibnamefont {Wzietek}}, \bibinfo {author} {\bibfnamefont
  {P.}~\bibnamefont {Metcalf}}, \ and\ \bibinfo {author} {\bibfnamefont
  {M.}~\bibnamefont {Marsi}},\ }\href {\doibase 10.1103/PhysRevLett.104.047401}
  {\bibfield  {journal} {\bibinfo  {journal} {Phys. Rev. Lett.}\ }\textbf
  {\bibinfo {volume} {104}},\ \bibinfo {pages} {047401} (\bibinfo {year}
  {2010})}\BibitemShut {NoStop}%
\bibitem [{\citenamefont {Homm}\ \emph {et~al.}(2015)\citenamefont {Homm},
  \citenamefont {Dillemans}, \citenamefont {Menghini}, \citenamefont
  {Van~Bilzen}, \citenamefont {Bakalov}, \citenamefont {Su}, \citenamefont
  {Lieten}, \citenamefont {Houssa}, \citenamefont {Nasr~Esfahani},
  \citenamefont {Covaci}, \citenamefont {Peeters}, \citenamefont {Seo},\ and\
  \citenamefont {Locquet}}]{Pia2017Collapse}%
  \BibitemOpen
  \bibfield  {author} {\bibinfo {author} {\bibfnamefont {P.}~\bibnamefont
  {Homm}}, \bibinfo {author} {\bibfnamefont {L.}~\bibnamefont {Dillemans}},
  \bibinfo {author} {\bibfnamefont {M.}~\bibnamefont {Menghini}}, \bibinfo
  {author} {\bibfnamefont {B.}~\bibnamefont {Van~Bilzen}}, \bibinfo {author}
  {\bibfnamefont {P.}~\bibnamefont {Bakalov}}, \bibinfo {author} {\bibfnamefont
  {C.-Y.}\ \bibnamefont {Su}}, \bibinfo {author} {\bibfnamefont
  {R.}~\bibnamefont {Lieten}}, \bibinfo {author} {\bibfnamefont
  {M.}~\bibnamefont {Houssa}}, \bibinfo {author} {\bibfnamefont
  {D.}~\bibnamefont {Nasr~Esfahani}}, \bibinfo {author} {\bibfnamefont
  {L.}~\bibnamefont {Covaci}}, \bibinfo {author} {\bibfnamefont {F.~M.}\
  \bibnamefont {Peeters}}, \bibinfo {author} {\bibfnamefont {J.~W.}\
  \bibnamefont {Seo}}, \ and\ \bibinfo {author} {\bibfnamefont {J.-P.}\
  \bibnamefont {Locquet}},\ }\href {\doibase 10.1063/1.4931372} {\bibfield
  {journal} {\bibinfo  {journal} {Applied Physics Letters}\ }\textbf {\bibinfo
  {volume} {107}},\ \bibinfo {pages} {111904} (\bibinfo {year}
  {2015})}\BibitemShut {NoStop}%
\bibitem [{\citenamefont {Homm}\ \emph {et~al.}(2021)\citenamefont {Homm},
  \citenamefont {Menghini}, \citenamefont {Seo}, \citenamefont {Peters},\ and\
  \citenamefont {Locquet}}]{Pia2021APLMaterial}%
  \BibitemOpen
  \bibfield  {author} {\bibinfo {author} {\bibfnamefont {P.}~\bibnamefont
  {Homm}}, \bibinfo {author} {\bibfnamefont {M.}~\bibnamefont {Menghini}},
  \bibinfo {author} {\bibfnamefont {J.~W.}\ \bibnamefont {Seo}}, \bibinfo
  {author} {\bibfnamefont {S.}~\bibnamefont {Peters}}, \ and\ \bibinfo {author}
  {\bibfnamefont {J.~P.}\ \bibnamefont {Locquet}},\ }\href {\doibase
  10.1063/5.0035865} {\bibfield  {journal} {\bibinfo  {journal} {APL
  Materials}\ }\textbf {\bibinfo {volume} {9}},\ \bibinfo {pages} {021116}
  (\bibinfo {year} {2021})}\BibitemShut {NoStop}%
\bibitem [{\citenamefont {Grieger}\ and\ \citenamefont
  {Lechermann}(2014)}]{Grieger2014EffectCrdoping}%
  \BibitemOpen
  \bibfield  {author} {\bibinfo {author} {\bibfnamefont {D.}~\bibnamefont
  {Grieger}}\ and\ \bibinfo {author} {\bibfnamefont {F.}~\bibnamefont
  {Lechermann}},\ }\href {\doibase 10.1103/PhysRevB.90.115115} {\bibfield
  {journal} {\bibinfo  {journal} {Phys. Rev. B}\ }\textbf {\bibinfo {volume}
  {90}},\ \bibinfo {pages} {115115} (\bibinfo {year} {2014})}\BibitemShut
  {NoStop}%
\bibitem [{\citenamefont {Poteryaev}\ \emph {et~al.}(2007)\citenamefont
  {Poteryaev}, \citenamefont {Tomczak}, \citenamefont {Biermann}, \citenamefont
  {Georges}, \citenamefont {Lichtenstein}, \citenamefont {Rubtsov},
  \citenamefont {Saha-Dasgupta},\ and\ \citenamefont
  {Andersen}}]{Poteryaev2007EnhancedCF}%
  \BibitemOpen
  \bibfield  {author} {\bibinfo {author} {\bibfnamefont {A.~I.}\ \bibnamefont
  {Poteryaev}}, \bibinfo {author} {\bibfnamefont {J.~M.}\ \bibnamefont
  {Tomczak}}, \bibinfo {author} {\bibfnamefont {S.}~\bibnamefont {Biermann}},
  \bibinfo {author} {\bibfnamefont {A.}~\bibnamefont {Georges}}, \bibinfo
  {author} {\bibfnamefont {A.~I.}\ \bibnamefont {Lichtenstein}}, \bibinfo
  {author} {\bibfnamefont {A.~N.}\ \bibnamefont {Rubtsov}}, \bibinfo {author}
  {\bibfnamefont {T.}~\bibnamefont {Saha-Dasgupta}}, \ and\ \bibinfo {author}
  {\bibfnamefont {O.~K.}\ \bibnamefont {Andersen}},\ }\href {\doibase
  10.1103/PhysRevB.76.085127} {\bibfield  {journal} {\bibinfo  {journal} {Phys.
  Rev. B}\ }\textbf {\bibinfo {volume} {76}},\ \bibinfo {pages} {085127}
  (\bibinfo {year} {2007})}\BibitemShut {NoStop}%
\bibitem [{\citenamefont {Pickem}\ \emph {et~al.}(2021)\citenamefont {Pickem},
  \citenamefont {Kaufmann}, \citenamefont {Held},\ and\ \citenamefont
  {Tomczak}}]{Zoology2021}%
  \BibitemOpen
  \bibfield  {author} {\bibinfo {author} {\bibfnamefont {M.}~\bibnamefont
  {Pickem}}, \bibinfo {author} {\bibfnamefont {J.}~\bibnamefont {Kaufmann}},
  \bibinfo {author} {\bibfnamefont {K.}~\bibnamefont {Held}}, \ and\ \bibinfo
  {author} {\bibfnamefont {J.~M.}\ \bibnamefont {Tomczak}},\ }\href {\doibase
  10.1103/PhysRevB.104.024307} {\bibfield  {journal} {\bibinfo  {journal}
  {Phys. Rev. B}\ }\textbf {\bibinfo {volume} {104}},\ \bibinfo {pages}
  {024307} (\bibinfo {year} {2021})}\BibitemShut {NoStop}%
\bibitem [{\citenamefont {Klebel-Knobloch}\ \emph {et~al.}(2021)\citenamefont
  {Klebel-Knobloch}, \citenamefont {Sch\"afer}, \citenamefont {Toschi},\ and\
  \citenamefont {Tomczak}}]{AnisotropyCorrelations2021}%
  \BibitemOpen
  \bibfield  {author} {\bibinfo {author} {\bibfnamefont {B.}~\bibnamefont
  {Klebel-Knobloch}}, \bibinfo {author} {\bibfnamefont {T.}~\bibnamefont
  {Sch\"afer}}, \bibinfo {author} {\bibfnamefont {A.}~\bibnamefont {Toschi}}, \
  and\ \bibinfo {author} {\bibfnamefont {J.~M.}\ \bibnamefont {Tomczak}},\
  }\href {\doibase 10.1103/PhysRevB.103.045121} {\bibfield  {journal} {\bibinfo
   {journal} {Phys. Rev. B}\ }\textbf {\bibinfo {volume} {103}},\ \bibinfo
  {pages} {045121} (\bibinfo {year} {2021})}\BibitemShut {NoStop}%
\bibitem [{\citenamefont {Mellaerts}\ \emph {et~al.}(2021)\citenamefont
  {Mellaerts}, \citenamefont {Meng}, \citenamefont {Menghini}, \citenamefont
  {Afanasiev}, \citenamefont {Seo}, \citenamefont {Houssa},\ and\ \citenamefont
  {Locquet}}]{2DV2O3Simon}%
  \BibitemOpen
  \bibfield  {author} {\bibinfo {author} {\bibfnamefont {S.}~\bibnamefont
  {Mellaerts}}, \bibinfo {author} {\bibfnamefont {R.}~\bibnamefont {Meng}},
  \bibinfo {author} {\bibfnamefont {M.}~\bibnamefont {Menghini}}, \bibinfo
  {author} {\bibfnamefont {V.}~\bibnamefont {Afanasiev}}, \bibinfo {author}
  {\bibfnamefont {J.~W.}\ \bibnamefont {Seo}}, \bibinfo {author} {\bibfnamefont
  {M.}~\bibnamefont {Houssa}}, \ and\ \bibinfo {author} {\bibfnamefont {J.-P.}\
  \bibnamefont {Locquet}},\ }\href {\doibase 10.1038/s41699-021-00245-w}
  {\bibfield  {journal} {\bibinfo  {journal} {npj 2D Materials and
  Applications}\ }\textbf {\bibinfo {volume} {5}},\ \bibinfo {pages} {65}
  (\bibinfo {year} {2021})}\BibitemShut {NoStop}%
\bibitem [{\citenamefont {Ponchut}\ \emph {et~al.}(2011)\citenamefont
  {Ponchut}, \citenamefont {Rigal}, \citenamefont {Cl\'ement}, \citenamefont
  {Papillon}, \citenamefont {Homs},\ and\ \citenamefont
  {Petitdemange}}]{synchrotronXRD}%
  \BibitemOpen
  \bibfield  {author} {\bibinfo {author} {\bibfnamefont {C.}~\bibnamefont
  {Ponchut}}, \bibinfo {author} {\bibfnamefont {J.~M.}\ \bibnamefont {Rigal}},
  \bibinfo {author} {\bibfnamefont {J.}~\bibnamefont {Cl\'ement}}, \bibinfo
  {author} {\bibfnamefont {E.}~\bibnamefont {Papillon}}, \bibinfo {author}
  {\bibfnamefont {A.}~\bibnamefont {Homs}}, \ and\ \bibinfo {author}
  {\bibfnamefont {S.}~\bibnamefont {Petitdemange}},\ }\href {\doibase
  10.1088/1748-0221/6/01/C01069} {\bibfield  {journal} {\bibinfo  {journal}
  {Journal of Instrumentation}\ }\textbf {\bibinfo {volume} {6}},\ \bibinfo
  {pages} {C01069} (\bibinfo {year} {2011})}\BibitemShut {NoStop}%
\bibitem [{\citenamefont {Wriedt}(1989)}]{Wriedt1989}%
  \BibitemOpen
  \bibfield  {author} {\bibinfo {author} {\bibfnamefont {H.~A.}\ \bibnamefont
  {Wriedt}},\ }\href {\doibase 10.1007/BF02877512} {\bibfield  {journal}
  {\bibinfo  {journal} {Bulletin of Alloy Phase Diagrams}\ }\textbf {\bibinfo
  {volume} {10}},\ \bibinfo {pages} {271} (\bibinfo {year} {1989})}\BibitemShut
  {NoStop}%
\bibitem [{\citenamefont {Bahlawane}\ and\ \citenamefont
  {Lenoble}(2014)}]{Bahlawane2014}%
  \BibitemOpen
  \bibfield  {author} {\bibinfo {author} {\bibfnamefont {N.}~\bibnamefont
  {Bahlawane}}\ and\ \bibinfo {author} {\bibfnamefont {D.}~\bibnamefont
  {Lenoble}},\ }\href {\doibase https://doi.org/10.1002/cvde.201400057}
  {\bibfield  {journal} {\bibinfo  {journal} {Chemical Vapor Deposition}\
  }\textbf {\bibinfo {volume} {20}},\ \bibinfo {pages} {299} (\bibinfo {year}
  {2014})}\BibitemShut {NoStop}%
\bibitem [{\citenamefont {Silversmit}\ \emph {et~al.}(2004)\citenamefont
  {Silversmit}, \citenamefont {Depla}, \citenamefont {Poelman}, \citenamefont
  {Marin},\ and\ \citenamefont {{De Gryse}}}]{SILVERSMIT2004167}%
  \BibitemOpen
  \bibfield  {author} {\bibinfo {author} {\bibfnamefont {G.}~\bibnamefont
  {Silversmit}}, \bibinfo {author} {\bibfnamefont {D.}~\bibnamefont {Depla}},
  \bibinfo {author} {\bibfnamefont {H.}~\bibnamefont {Poelman}}, \bibinfo
  {author} {\bibfnamefont {G.~B.}\ \bibnamefont {Marin}}, \ and\ \bibinfo
  {author} {\bibfnamefont {R.}~\bibnamefont {{De Gryse}}},\ }\href {\doibase
  https://doi.org/10.1016/j.elspec.2004.03.004} {\bibfield  {journal} {\bibinfo
   {journal} {Journal of Electron Spectroscopy and Related Phenomena}\ }\textbf
  {\bibinfo {volume} {135}},\ \bibinfo {pages} {167} (\bibinfo {year}
  {2004})}\BibitemShut {NoStop}%
\bibitem [{\citenamefont {Sawatzky}\ and\ \citenamefont
  {Post}(1979)}]{Sawatzky1979}%
  \BibitemOpen
  \bibfield  {author} {\bibinfo {author} {\bibfnamefont {G.~A.}\ \bibnamefont
  {Sawatzky}}\ and\ \bibinfo {author} {\bibfnamefont {D.}~\bibnamefont
  {Post}},\ }\href {\doibase 10.1103/PhysRevB.20.1546} {\bibfield  {journal}
  {\bibinfo  {journal} {Phys. Rev. B}\ }\textbf {\bibinfo {volume} {20}},\
  \bibinfo {pages} {1546} (\bibinfo {year} {1979})}\BibitemShut {NoStop}%
\bibitem [{\citenamefont {Luo}\ \emph {et~al.}(2004)\citenamefont {Luo},
  \citenamefont {Guo},\ and\ \citenamefont {Wang}}]{Luo2004}%
  \BibitemOpen
  \bibfield  {author} {\bibinfo {author} {\bibfnamefont {Q.}~\bibnamefont
  {Luo}}, \bibinfo {author} {\bibfnamefont {Q.}~\bibnamefont {Guo}}, \ and\
  \bibinfo {author} {\bibfnamefont {E.~G.}\ \bibnamefont {Wang}},\ }\href
  {\doibase 10.1063/1.1690107} {\bibfield  {journal} {\bibinfo  {journal}
  {Applied Physics Letters}\ }\textbf {\bibinfo {volume} {84}},\ \bibinfo
  {pages} {2337} (\bibinfo {year} {2004})}\BibitemShut {NoStop}%
\bibitem [{\citenamefont {Polewczyk}\ \emph {et~al.}(2023)\citenamefont
  {Polewczyk}, \citenamefont {Chaluvadi}, \citenamefont {Dagur}, \citenamefont
  {Mazzola}, \citenamefont {{Punathum Chalil}}, \citenamefont {Petrov},
  \citenamefont {Fujii}, \citenamefont {Panaccione}, \citenamefont {Rossi},
  \citenamefont {Orgiani}, \citenamefont {Vinai},\ and\ \citenamefont
  {Torelli}}]{POLEWCZYK2023155462}%
  \BibitemOpen
  \bibfield  {author} {\bibinfo {author} {\bibfnamefont {V.}~\bibnamefont
  {Polewczyk}}, \bibinfo {author} {\bibfnamefont {S.}~\bibnamefont
  {Chaluvadi}}, \bibinfo {author} {\bibfnamefont {D.}~\bibnamefont {Dagur}},
  \bibinfo {author} {\bibfnamefont {F.}~\bibnamefont {Mazzola}}, \bibinfo
  {author} {\bibfnamefont {S.}~\bibnamefont {{Punathum Chalil}}}, \bibinfo
  {author} {\bibfnamefont {A.}~\bibnamefont {Petrov}}, \bibinfo {author}
  {\bibfnamefont {J.}~\bibnamefont {Fujii}}, \bibinfo {author} {\bibfnamefont
  {G.}~\bibnamefont {Panaccione}}, \bibinfo {author} {\bibfnamefont
  {G.}~\bibnamefont {Rossi}}, \bibinfo {author} {\bibfnamefont
  {P.}~\bibnamefont {Orgiani}}, \bibinfo {author} {\bibfnamefont
  {G.}~\bibnamefont {Vinai}}, \ and\ \bibinfo {author} {\bibfnamefont
  {P.}~\bibnamefont {Torelli}},\ }\href {\doibase
  https://doi.org/10.1016/j.apsusc.2022.155462} {\bibfield  {journal} {\bibinfo
   {journal} {Applied Surface Science}\ }\textbf {\bibinfo {volume} {610}},\
  \bibinfo {pages} {155462} (\bibinfo {year} {2023})}\BibitemShut {NoStop}%
\bibitem [{\citenamefont {Dillemans}\ \emph {et~al.}(2014)\citenamefont
  {Dillemans}, \citenamefont {Smets}, \citenamefont {Lieten}, \citenamefont
  {Menghini}, \citenamefont {Su},\ and\ \citenamefont
  {Locquet}}]{Dillemans2014}%
  \BibitemOpen
  \bibfield  {author} {\bibinfo {author} {\bibfnamefont {L.}~\bibnamefont
  {Dillemans}}, \bibinfo {author} {\bibfnamefont {T.}~\bibnamefont {Smets}},
  \bibinfo {author} {\bibfnamefont {R.~R.}\ \bibnamefont {Lieten}}, \bibinfo
  {author} {\bibfnamefont {M.}~\bibnamefont {Menghini}}, \bibinfo {author}
  {\bibfnamefont {C.-Y.}\ \bibnamefont {Su}}, \ and\ \bibinfo {author}
  {\bibfnamefont {J.-P.}\ \bibnamefont {Locquet}},\ }\href {\doibase
  10.1063/1.4866004} {\bibfield  {journal} {\bibinfo  {journal} {Applied
  Physics Letters}\ }\textbf {\bibinfo {volume} {104}} (\bibinfo {year}
  {2014}),\ 10.1063/1.4866004}\BibitemShut {NoStop}%
\bibitem [{\citenamefont {Imada}\ \emph {et~al.}(1998)\citenamefont {Imada},
  \citenamefont {Fujimori},\ and\ \citenamefont {Tokura}}]{Imada1998}%
  \BibitemOpen
  \bibfield  {author} {\bibinfo {author} {\bibfnamefont {M.}~\bibnamefont
  {Imada}}, \bibinfo {author} {\bibfnamefont {A.}~\bibnamefont {Fujimori}}, \
  and\ \bibinfo {author} {\bibfnamefont {Y.}~\bibnamefont {Tokura}},\ }\href
  {\doibase 10.1103/RevModPhys.70.1039} {\bibfield  {journal} {\bibinfo
  {journal} {Rev. Mod. Phys.}\ }\textbf {\bibinfo {volume} {70}},\ \bibinfo
  {pages} {1039} (\bibinfo {year} {1998})}\BibitemShut {NoStop}%
\bibitem [{\citenamefont {Held}\ \emph {et~al.}(2001)\citenamefont {Held},
  \citenamefont {Keller}, \citenamefont {Eyert}, \citenamefont {Vollhardt},\
  and\ \citenamefont {Anisimov}}]{Held2001}%
  \BibitemOpen
  \bibfield  {author} {\bibinfo {author} {\bibfnamefont {K.}~\bibnamefont
  {Held}}, \bibinfo {author} {\bibfnamefont {G.}~\bibnamefont {Keller}},
  \bibinfo {author} {\bibfnamefont {V.}~\bibnamefont {Eyert}}, \bibinfo
  {author} {\bibfnamefont {D.}~\bibnamefont {Vollhardt}}, \ and\ \bibinfo
  {author} {\bibfnamefont {V.~I.}\ \bibnamefont {Anisimov}},\ }\href {\doibase
  10.1103/PhysRevLett.86.5345} {\bibfield  {journal} {\bibinfo  {journal}
  {Phys. Rev. Lett.}\ }\textbf {\bibinfo {volume} {86}},\ \bibinfo {pages}
  {5345} (\bibinfo {year} {2001})}\BibitemShut {NoStop}%
\bibitem [{\citenamefont {Hansmann}\ \emph {et~al.}(2013)\citenamefont
  {Hansmann}, \citenamefont {Toschi}, \citenamefont {Sangiovanni},
  \citenamefont {Saha-Dasgupta}, \citenamefont {Lupi}, \citenamefont {Marsi},\
  and\ \citenamefont {Held}}]{Hansmann2013}%
  \BibitemOpen
  \bibfield  {author} {\bibinfo {author} {\bibfnamefont {P.}~\bibnamefont
  {Hansmann}}, \bibinfo {author} {\bibfnamefont {A.}~\bibnamefont {Toschi}},
  \bibinfo {author} {\bibfnamefont {G.}~\bibnamefont {Sangiovanni}}, \bibinfo
  {author} {\bibfnamefont {T.}~\bibnamefont {Saha-Dasgupta}}, \bibinfo {author}
  {\bibfnamefont {S.}~\bibnamefont {Lupi}}, \bibinfo {author} {\bibfnamefont
  {M.}~\bibnamefont {Marsi}}, \ and\ \bibinfo {author} {\bibfnamefont
  {K.}~\bibnamefont {Held}},\ }\href {\doibase
  https://doi.org/10.1002/pssb.201248476} {\bibfield  {journal} {\bibinfo
  {journal} {physica status solidi (b)}\ }\textbf {\bibinfo {volume} {250}},\
  \bibinfo {pages} {1251} (\bibinfo {year} {2013})}\BibitemShut {NoStop}%
\bibitem [{\citenamefont {Borghi}\ \emph {et~al.}(2009)\citenamefont {Borghi},
  \citenamefont {Fabrizio},\ and\ \citenamefont {Tosatti}}]{Borghi2009QP}%
  \BibitemOpen
  \bibfield  {author} {\bibinfo {author} {\bibfnamefont {G.}~\bibnamefont
  {Borghi}}, \bibinfo {author} {\bibfnamefont {M.}~\bibnamefont {Fabrizio}}, \
  and\ \bibinfo {author} {\bibfnamefont {E.}~\bibnamefont {Tosatti}},\ }\href
  {\doibase 10.1103/PhysRevLett.102.066806} {\bibfield  {journal} {\bibinfo
  {journal} {Phys. Rev. Lett.}\ }\textbf {\bibinfo {volume} {102}},\ \bibinfo
  {pages} {066806} (\bibinfo {year} {2009})}\BibitemShut {NoStop}%
\bibitem [{\citenamefont {Rodolakis}\ \emph {et~al.}(2009)\citenamefont
  {Rodolakis}, \citenamefont {Mansart}, \citenamefont {Papalazarou},
  \citenamefont {Gorovikov}, \citenamefont {Vilmercati}, \citenamefont
  {Petaccia}, \citenamefont {Goldoni}, \citenamefont {Rueff}, \citenamefont
  {Lupi}, \citenamefont {Metcalf},\ and\ \citenamefont
  {Marsi}}]{Rodolakis2009QP}%
  \BibitemOpen
  \bibfield  {author} {\bibinfo {author} {\bibfnamefont {F.}~\bibnamefont
  {Rodolakis}}, \bibinfo {author} {\bibfnamefont {B.}~\bibnamefont {Mansart}},
  \bibinfo {author} {\bibfnamefont {E.}~\bibnamefont {Papalazarou}}, \bibinfo
  {author} {\bibfnamefont {S.}~\bibnamefont {Gorovikov}}, \bibinfo {author}
  {\bibfnamefont {P.}~\bibnamefont {Vilmercati}}, \bibinfo {author}
  {\bibfnamefont {L.}~\bibnamefont {Petaccia}}, \bibinfo {author}
  {\bibfnamefont {A.}~\bibnamefont {Goldoni}}, \bibinfo {author} {\bibfnamefont
  {J.~P.}\ \bibnamefont {Rueff}}, \bibinfo {author} {\bibfnamefont
  {S.}~\bibnamefont {Lupi}}, \bibinfo {author} {\bibfnamefont {P.}~\bibnamefont
  {Metcalf}}, \ and\ \bibinfo {author} {\bibfnamefont {M.}~\bibnamefont
  {Marsi}},\ }\href {\doibase 10.1103/PhysRevLett.102.066805} {\bibfield
  {journal} {\bibinfo  {journal} {Phys. Rev. Lett.}\ }\textbf {\bibinfo
  {volume} {102}},\ \bibinfo {pages} {066805} (\bibinfo {year}
  {2009})}\BibitemShut {NoStop}%
\bibitem [{\citenamefont {Georges}\ \emph {et~al.}(1996)\citenamefont
  {Georges}, \citenamefont {Kotliar}, \citenamefont {Krauth},\ and\
  \citenamefont {Rozenberg}}]{Georges1996}%
  \BibitemOpen
  \bibfield  {author} {\bibinfo {author} {\bibfnamefont {A.}~\bibnamefont
  {Georges}}, \bibinfo {author} {\bibfnamefont {G.}~\bibnamefont {Kotliar}},
  \bibinfo {author} {\bibfnamefont {W.}~\bibnamefont {Krauth}}, \ and\ \bibinfo
  {author} {\bibfnamefont {M.~J.}\ \bibnamefont {Rozenberg}},\ }\href {\doibase
  10.1103/RevModPhys.68.13} {\bibfield  {journal} {\bibinfo  {journal} {Rev.
  Mod. Phys.}\ }\textbf {\bibinfo {volume} {68}},\ \bibinfo {pages} {13}
  (\bibinfo {year} {1996})}\BibitemShut {NoStop}%
\bibitem [{\citenamefont {Thees}\ \emph {et~al.}(2021)\citenamefont {Thees},
  \citenamefont {Lee}, \citenamefont {Bouwmeester}, \citenamefont
  {Rezende-Gon{\c c}alves}, \citenamefont {David}, \citenamefont {Zimmers},
  \citenamefont {Fortuna}, \citenamefont {Frantzeskakis}, \citenamefont
  {Vargas}, \citenamefont {Kalcheim}, \citenamefont {F{\`e}vre}, \citenamefont
  {Horiba}, \citenamefont {Kumigashira}, \citenamefont {Biermann},
  \citenamefont {Trastoy}, \citenamefont {Rozenberg}, \citenamefont
  {Schuller},\ and\ \citenamefont {Santander-Syro}}]{Thees2021ARPES}%
  \BibitemOpen
  \bibfield  {author} {\bibinfo {author} {\bibfnamefont {M.}~\bibnamefont
  {Thees}}, \bibinfo {author} {\bibfnamefont {M.-H.}\ \bibnamefont {Lee}},
  \bibinfo {author} {\bibfnamefont {R.~L.}\ \bibnamefont {Bouwmeester}},
  \bibinfo {author} {\bibfnamefont {P.~H.}\ \bibnamefont {Rezende-Gon{\c
  c}alves}}, \bibinfo {author} {\bibfnamefont {E.}~\bibnamefont {David}},
  \bibinfo {author} {\bibfnamefont {A.}~\bibnamefont {Zimmers}}, \bibinfo
  {author} {\bibfnamefont {F.}~\bibnamefont {Fortuna}}, \bibinfo {author}
  {\bibfnamefont {E.}~\bibnamefont {Frantzeskakis}}, \bibinfo {author}
  {\bibfnamefont {N.~M.}\ \bibnamefont {Vargas}}, \bibinfo {author}
  {\bibfnamefont {Y.}~\bibnamefont {Kalcheim}}, \bibinfo {author}
  {\bibfnamefont {P.~L.}\ \bibnamefont {F{\`e}vre}}, \bibinfo {author}
  {\bibfnamefont {K.}~\bibnamefont {Horiba}}, \bibinfo {author} {\bibfnamefont
  {H.}~\bibnamefont {Kumigashira}}, \bibinfo {author} {\bibfnamefont
  {S.}~\bibnamefont {Biermann}}, \bibinfo {author} {\bibfnamefont
  {J.}~\bibnamefont {Trastoy}}, \bibinfo {author} {\bibfnamefont {M.~J.}\
  \bibnamefont {Rozenberg}}, \bibinfo {author} {\bibfnamefont {I.~K.}\
  \bibnamefont {Schuller}}, \ and\ \bibinfo {author} {\bibfnamefont {A.~F.}\
  \bibnamefont {Santander-Syro}},\ }\href {\doibase 10.1126/sciadv.abj1164}
  {\bibfield  {journal} {\bibinfo  {journal} {Science Advances}\ }\textbf
  {\bibinfo {volume} {7}},\ \bibinfo {pages} {eabj1164} (\bibinfo {year}
  {2021})}\BibitemShut {NoStop}%
\bibitem [{\citenamefont {Park}\ \emph {et~al.}(2000)\citenamefont {Park},
  \citenamefont {Tjeng}, \citenamefont {Tanaka}, \citenamefont {Allen},
  \citenamefont {Chen}, \citenamefont {Metcalf}, \citenamefont {Honig},
  \citenamefont {de~Groot},\ and\ \citenamefont
  {Sawatzky}}]{Park2000occupation}%
  \BibitemOpen
  \bibfield  {author} {\bibinfo {author} {\bibfnamefont {J.-H.}\ \bibnamefont
  {Park}}, \bibinfo {author} {\bibfnamefont {L.~H.}\ \bibnamefont {Tjeng}},
  \bibinfo {author} {\bibfnamefont {A.}~\bibnamefont {Tanaka}}, \bibinfo
  {author} {\bibfnamefont {J.~W.}\ \bibnamefont {Allen}}, \bibinfo {author}
  {\bibfnamefont {C.~T.}\ \bibnamefont {Chen}}, \bibinfo {author}
  {\bibfnamefont {P.}~\bibnamefont {Metcalf}}, \bibinfo {author} {\bibfnamefont
  {J.~M.}\ \bibnamefont {Honig}}, \bibinfo {author} {\bibfnamefont {F.~M.~F.}\
  \bibnamefont {de~Groot}}, \ and\ \bibinfo {author} {\bibfnamefont {G.~A.}\
  \bibnamefont {Sawatzky}},\ }\href {\doibase 10.1103/PhysRevB.61.11506}
  {\bibfield  {journal} {\bibinfo  {journal} {Phys. Rev. B}\ }\textbf {\bibinfo
  {volume} {61}},\ \bibinfo {pages} {11506} (\bibinfo {year}
  {2000})}\BibitemShut {NoStop}%
\bibitem [{\citenamefont {Lo~Vecchio}\ \emph {et~al.}(2015)\citenamefont
  {Lo~Vecchio}, \citenamefont {Baldassarre}, \citenamefont {D'Apuzzo},
  \citenamefont {Limaj}, \citenamefont {Nicoletti}, \citenamefont {Perucchi},
  \citenamefont {Fan}, \citenamefont {Metcalf}, \citenamefont {Marsi},\ and\
  \citenamefont {Lupi}}]{Vecchio2015Optics}%
  \BibitemOpen
  \bibfield  {author} {\bibinfo {author} {\bibfnamefont {I.}~\bibnamefont
  {Lo~Vecchio}}, \bibinfo {author} {\bibfnamefont {L.}~\bibnamefont
  {Baldassarre}}, \bibinfo {author} {\bibfnamefont {F.}~\bibnamefont
  {D'Apuzzo}}, \bibinfo {author} {\bibfnamefont {O.}~\bibnamefont {Limaj}},
  \bibinfo {author} {\bibfnamefont {D.}~\bibnamefont {Nicoletti}}, \bibinfo
  {author} {\bibfnamefont {A.}~\bibnamefont {Perucchi}}, \bibinfo {author}
  {\bibfnamefont {L.}~\bibnamefont {Fan}}, \bibinfo {author} {\bibfnamefont
  {P.}~\bibnamefont {Metcalf}}, \bibinfo {author} {\bibfnamefont
  {M.}~\bibnamefont {Marsi}}, \ and\ \bibinfo {author} {\bibfnamefont
  {S.}~\bibnamefont {Lupi}},\ }\href {\doibase 10.1103/PhysRevB.91.155133}
  {\bibfield  {journal} {\bibinfo  {journal} {Phys. Rev. B}\ }\textbf {\bibinfo
  {volume} {91}},\ \bibinfo {pages} {155133} (\bibinfo {year}
  {2015})}\BibitemShut {NoStop}%
\bibitem [{\citenamefont {Yoshimatsu}\ \emph {et~al.}(2010)\citenamefont
  {Yoshimatsu}, \citenamefont {Okabe}, \citenamefont {Kumigashira},
  \citenamefont {Okamoto}, \citenamefont {Aizaki}, \citenamefont {Fujimori},\
  and\ \citenamefont {Oshima}}]{Yoshimatsu2010SrVO3}%
  \BibitemOpen
  \bibfield  {author} {\bibinfo {author} {\bibfnamefont {K.}~\bibnamefont
  {Yoshimatsu}}, \bibinfo {author} {\bibfnamefont {T.}~\bibnamefont {Okabe}},
  \bibinfo {author} {\bibfnamefont {H.}~\bibnamefont {Kumigashira}}, \bibinfo
  {author} {\bibfnamefont {S.}~\bibnamefont {Okamoto}}, \bibinfo {author}
  {\bibfnamefont {S.}~\bibnamefont {Aizaki}}, \bibinfo {author} {\bibfnamefont
  {A.}~\bibnamefont {Fujimori}}, \ and\ \bibinfo {author} {\bibfnamefont
  {M.}~\bibnamefont {Oshima}},\ }\href {\doibase
  10.1103/PhysRevLett.104.147601} {\bibfield  {journal} {\bibinfo  {journal}
  {Phys. Rev. Lett.}\ }\textbf {\bibinfo {volume} {104}},\ \bibinfo {pages}
  {147601} (\bibinfo {year} {2010})}\BibitemShut {NoStop}%
\bibitem [{\citenamefont {Gu}\ \emph {et~al.}(2013)\citenamefont {Gu},
  \citenamefont {Laverock}, \citenamefont {Chen}, \citenamefont {Smith},
  \citenamefont {Wolf},\ and\ \citenamefont {Lu}}]{Gu2013CaVO3}%
  \BibitemOpen
  \bibfield  {author} {\bibinfo {author} {\bibfnamefont {M.}~\bibnamefont
  {Gu}}, \bibinfo {author} {\bibfnamefont {J.}~\bibnamefont {Laverock}},
  \bibinfo {author} {\bibfnamefont {B.}~\bibnamefont {Chen}}, \bibinfo {author}
  {\bibfnamefont {K.~E.}\ \bibnamefont {Smith}}, \bibinfo {author}
  {\bibfnamefont {S.~A.}\ \bibnamefont {Wolf}}, \ and\ \bibinfo {author}
  {\bibfnamefont {J.}~\bibnamefont {Lu}},\ }\href {\doibase 10.1063/1.4798963}
  {\bibfield  {journal} {\bibinfo  {journal} {Journal of Applied Physics}\
  }\textbf {\bibinfo {volume} {113}},\ \bibinfo {pages} {133704} (\bibinfo
  {year} {2013})}\BibitemShut {NoStop}%
\bibitem [{\citenamefont {Sakai}\ \emph {et~al.}(2013)\citenamefont {Sakai},
  \citenamefont {Tamamitsu}, \citenamefont {Yoshimatsu}, \citenamefont
  {Okamoto}, \citenamefont {Horiba}, \citenamefont {Oshima},\ and\
  \citenamefont {Kumigashira}}]{Sakai2013}%
  \BibitemOpen
  \bibfield  {author} {\bibinfo {author} {\bibfnamefont {E.}~\bibnamefont
  {Sakai}}, \bibinfo {author} {\bibfnamefont {M.}~\bibnamefont {Tamamitsu}},
  \bibinfo {author} {\bibfnamefont {K.}~\bibnamefont {Yoshimatsu}}, \bibinfo
  {author} {\bibfnamefont {S.}~\bibnamefont {Okamoto}}, \bibinfo {author}
  {\bibfnamefont {K.}~\bibnamefont {Horiba}}, \bibinfo {author} {\bibfnamefont
  {M.}~\bibnamefont {Oshima}}, \ and\ \bibinfo {author} {\bibfnamefont
  {H.}~\bibnamefont {Kumigashira}},\ }\href {\doibase
  10.1103/PhysRevB.87.075132} {\bibfield  {journal} {\bibinfo  {journal} {Phys.
  Rev. B}\ }\textbf {\bibinfo {volume} {87}},\ \bibinfo {pages} {075132}
  (\bibinfo {year} {2013})}\BibitemShut {NoStop}%
\bibitem [{\citenamefont {Gu}\ \emph {et~al.}(2014)\citenamefont {Gu},
  \citenamefont {Wolf},\ and\ \citenamefont {Lu}}]{Gu2014}%
  \BibitemOpen
  \bibfield  {author} {\bibinfo {author} {\bibfnamefont {M.}~\bibnamefont
  {Gu}}, \bibinfo {author} {\bibfnamefont {S.~A.}\ \bibnamefont {Wolf}}, \ and\
  \bibinfo {author} {\bibfnamefont {J.}~\bibnamefont {Lu}},\ }\href {\doibase
  https://doi.org/10.1002/admi.201300126} {\bibfield  {journal} {\bibinfo
  {journal} {Advanced Materials Interfaces}\ }\textbf {\bibinfo {volume} {1}},\
  \bibinfo {pages} {1300126} (\bibinfo {year} {2014})}\BibitemShut {NoStop}%
\bibitem [{\citenamefont {Shiga}\ \emph {et~al.}(2020)\citenamefont {Shiga},
  \citenamefont {Yang}, \citenamefont {Hasegawa}, \citenamefont {Kanda},
  \citenamefont {Tokunaga}, \citenamefont {Yoshimatsu}, \citenamefont {Yukawa},
  \citenamefont {Kitamura}, \citenamefont {Horiba},\ and\ \citenamefont
  {Kumigashira}}]{Shiga2020}%
  \BibitemOpen
  \bibfield  {author} {\bibinfo {author} {\bibfnamefont {D.}~\bibnamefont
  {Shiga}}, \bibinfo {author} {\bibfnamefont {B.~E.}\ \bibnamefont {Yang}},
  \bibinfo {author} {\bibfnamefont {N.}~\bibnamefont {Hasegawa}}, \bibinfo
  {author} {\bibfnamefont {T.}~\bibnamefont {Kanda}}, \bibinfo {author}
  {\bibfnamefont {R.}~\bibnamefont {Tokunaga}}, \bibinfo {author}
  {\bibfnamefont {K.}~\bibnamefont {Yoshimatsu}}, \bibinfo {author}
  {\bibfnamefont {R.}~\bibnamefont {Yukawa}}, \bibinfo {author} {\bibfnamefont
  {M.}~\bibnamefont {Kitamura}}, \bibinfo {author} {\bibfnamefont
  {K.}~\bibnamefont {Horiba}}, \ and\ \bibinfo {author} {\bibfnamefont
  {H.}~\bibnamefont {Kumigashira}},\ }\href {\doibase
  10.1103/PhysRevB.102.115114} {\bibfield  {journal} {\bibinfo  {journal}
  {Phys. Rev. B}\ }\textbf {\bibinfo {volume} {102}},\ \bibinfo {pages}
  {115114} (\bibinfo {year} {2020})}\BibitemShut {NoStop}%
\bibitem [{\citenamefont {Kuroda}\ and\ \citenamefont
  {Fan}(1977)}]{Kuroda1977}%
  \BibitemOpen
  \bibfield  {author} {\bibinfo {author} {\bibfnamefont {N.}~\bibnamefont
  {Kuroda}}\ and\ \bibinfo {author} {\bibfnamefont {H.~Y.}\ \bibnamefont
  {Fan}},\ }\href {\doibase 10.1103/PhysRevB.16.5003} {\bibfield  {journal}
  {\bibinfo  {journal} {Phys. Rev. B}\ }\textbf {\bibinfo {volume} {16}},\
  \bibinfo {pages} {5003} (\bibinfo {year} {1977})}\BibitemShut {NoStop}%
\bibitem [{\citenamefont {Hsu}\ \emph {et~al.}(2023)\citenamefont {Hsu},
  \citenamefont {Mellaerts}, \citenamefont {Bellani}, \citenamefont {Homm},
  \citenamefont {Uchida}, \citenamefont {Menghini}, \citenamefont {Houssa},
  \citenamefont {Seo},\ and\ \citenamefont {Locquet}}]{2023RamanV2O3}%
  \BibitemOpen
  \bibfield  {author} {\bibinfo {author} {\bibfnamefont {W.-F.}\ \bibnamefont
  {Hsu}}, \bibinfo {author} {\bibfnamefont {S.}~\bibnamefont {Mellaerts}},
  \bibinfo {author} {\bibfnamefont {C.}~\bibnamefont {Bellani}}, \bibinfo
  {author} {\bibfnamefont {P.}~\bibnamefont {Homm}}, \bibinfo {author}
  {\bibfnamefont {N.}~\bibnamefont {Uchida}}, \bibinfo {author} {\bibfnamefont
  {M.}~\bibnamefont {Menghini}}, \bibinfo {author} {\bibfnamefont
  {M.}~\bibnamefont {Houssa}}, \bibinfo {author} {\bibfnamefont {J.~W.}\
  \bibnamefont {Seo}}, \ and\ \bibinfo {author} {\bibfnamefont {J.-P.}\
  \bibnamefont {Locquet}},\ }\href {\doibase 10.1103/PhysRevMaterials.7.074606}
  {\bibfield  {journal} {\bibinfo  {journal} {Phys. Rev. Mater.}\ }\textbf
  {\bibinfo {volume} {7}},\ \bibinfo {pages} {074606} (\bibinfo {year}
  {2023})}\BibitemShut {NoStop}%
\bibitem [{\citenamefont {Yang}\ and\ \citenamefont {Sladek}(1985)}]{Yang1985}%
  \BibitemOpen
  \bibfield  {author} {\bibinfo {author} {\bibfnamefont {H.}~\bibnamefont
  {Yang}}\ and\ \bibinfo {author} {\bibfnamefont {R.~J.}\ \bibnamefont
  {Sladek}},\ }\href {\doibase 10.1103/PhysRevB.32.6634} {\bibfield  {journal}
  {\bibinfo  {journal} {Phys. Rev. B}\ }\textbf {\bibinfo {volume} {32}},\
  \bibinfo {pages} {6634} (\bibinfo {year} {1985})}\BibitemShut {NoStop}%
\bibitem [{\citenamefont {Senet}\ \emph {et~al.}(1995)\citenamefont {Senet},
  \citenamefont {Lambin},\ and\ \citenamefont {Lucas}}]{Senet1195}%
  \BibitemOpen
  \bibfield  {author} {\bibinfo {author} {\bibfnamefont {P.}~\bibnamefont
  {Senet}}, \bibinfo {author} {\bibfnamefont {P.}~\bibnamefont {Lambin}}, \
  and\ \bibinfo {author} {\bibfnamefont {A.~A.}\ \bibnamefont {Lucas}},\ }\href
  {\doibase 10.1103/PhysRevLett.74.570} {\bibfield  {journal} {\bibinfo
  {journal} {Phys. Rev. Lett.}\ }\textbf {\bibinfo {volume} {74}},\ \bibinfo
  {pages} {570} (\bibinfo {year} {1995})}\BibitemShut {NoStop}%
\bibitem [{\citenamefont {Premper}\ \emph {et~al.}(2020)\citenamefont
  {Premper}, \citenamefont {Schumann}, \citenamefont {Dhaka}, \citenamefont
  {Polzin}, \citenamefont {Kostov}, \citenamefont {Goian}, \citenamefont
  {Sander},\ and\ \citenamefont {Widdra}}]{Premper2020}%
  \BibitemOpen
  \bibfield  {author} {\bibinfo {author} {\bibfnamefont {J.}~\bibnamefont
  {Premper}}, \bibinfo {author} {\bibfnamefont {F.~O.}\ \bibnamefont
  {Schumann}}, \bibinfo {author} {\bibfnamefont {A.}~\bibnamefont {Dhaka}},
  \bibinfo {author} {\bibfnamefont {S.}~\bibnamefont {Polzin}}, \bibinfo
  {author} {\bibfnamefont {K.~L.}\ \bibnamefont {Kostov}}, \bibinfo {author}
  {\bibfnamefont {V.}~\bibnamefont {Goian}}, \bibinfo {author} {\bibfnamefont
  {D.}~\bibnamefont {Sander}}, \ and\ \bibinfo {author} {\bibfnamefont
  {W.}~\bibnamefont {Widdra}},\ }\href {\doibase
  https://doi.org/10.1002/pssb.201900650} {\bibfield  {journal} {\bibinfo
  {journal} {physica status solidi (b)}\ }\textbf {\bibinfo {volume} {257}},\
  \bibinfo {pages} {1900650} (\bibinfo {year} {2020})}\BibitemShut {NoStop}%
\bibitem [{\citenamefont {Schober}\ \emph {et~al.}(2020)\citenamefont
  {Schober}, \citenamefont {Fowlie}, \citenamefont {Guennou}, \citenamefont
  {Weber}, \citenamefont {Zhao}, \citenamefont {Iniguez}, \citenamefont
  {Gibert}, \citenamefont {Triscone},\ and\ \citenamefont
  {Kreisel}}]{2020Schober}%
  \BibitemOpen
  \bibfield  {author} {\bibinfo {author} {\bibfnamefont {A.}~\bibnamefont
  {Schober}}, \bibinfo {author} {\bibfnamefont {J.}~\bibnamefont {Fowlie}},
  \bibinfo {author} {\bibfnamefont {M.}~\bibnamefont {Guennou}}, \bibinfo
  {author} {\bibfnamefont {M.~C.}\ \bibnamefont {Weber}}, \bibinfo {author}
  {\bibfnamefont {H.}~\bibnamefont {Zhao}}, \bibinfo {author} {\bibfnamefont
  {J.}~\bibnamefont {Iniguez}}, \bibinfo {author} {\bibfnamefont
  {M.}~\bibnamefont {Gibert}}, \bibinfo {author} {\bibfnamefont {J.-M.}\
  \bibnamefont {Triscone}}, \ and\ \bibinfo {author} {\bibfnamefont
  {J.}~\bibnamefont {Kreisel}},\ }\href {\doibase 10.1063/5.0010233} {\bibfield
   {journal} {\bibinfo  {journal} {APL Materials}\ }\textbf {\bibinfo {volume}
  {8}},\ \bibinfo {pages} {061102} (\bibinfo {year} {2020})}\BibitemShut
  {NoStop}%
\bibitem [{\citenamefont {Lin}\ \emph {et~al.}(2022)\citenamefont {Lin},
  \citenamefont {Li}, \citenamefont {Yu}, \citenamefont {Chen}, \citenamefont
  {Attfield},\ and\ \citenamefont {Xing}}]{Lin2022ChemicalPreview}%
  \BibitemOpen
  \bibfield  {author} {\bibinfo {author} {\bibfnamefont {K.}~\bibnamefont
  {Lin}}, \bibinfo {author} {\bibfnamefont {Q.}~\bibnamefont {Li}}, \bibinfo
  {author} {\bibfnamefont {R.}~\bibnamefont {Yu}}, \bibinfo {author}
  {\bibfnamefont {J.}~\bibnamefont {Chen}}, \bibinfo {author} {\bibfnamefont
  {J.~P.}\ \bibnamefont {Attfield}}, \ and\ \bibinfo {author} {\bibfnamefont
  {X.}~\bibnamefont {Xing}},\ }\href {\doibase 10.1039/D1CS00563D} {\bibfield
  {journal} {\bibinfo  {journal} {Chem. Soc. Rev.}\ }\textbf {\bibinfo {volume}
  {51}},\ \bibinfo {pages} {5351} (\bibinfo {year} {2022})}\BibitemShut
  {NoStop}%
\bibitem [{\citenamefont {Sun}\ \emph {et~al.}(2020)\citenamefont {Sun},
  \citenamefont {Li}, \citenamefont {Zhu}, \citenamefont {Liu}, \citenamefont
  {Lin}, \citenamefont {Wang}, \citenamefont {Zhang}, \citenamefont {Gu},
  \citenamefont {Deng}, \citenamefont {Chen},\ and\ \citenamefont
  {Xing}}]{Sun2020NP-NanowiresPTO}%
  \BibitemOpen
  \bibfield  {author} {\bibinfo {author} {\bibfnamefont {J.}~\bibnamefont
  {Sun}}, \bibinfo {author} {\bibfnamefont {Q.}~\bibnamefont {Li}}, \bibinfo
  {author} {\bibfnamefont {H.}~\bibnamefont {Zhu}}, \bibinfo {author}
  {\bibfnamefont {Z.}~\bibnamefont {Liu}}, \bibinfo {author} {\bibfnamefont
  {K.}~\bibnamefont {Lin}}, \bibinfo {author} {\bibfnamefont {N.}~\bibnamefont
  {Wang}}, \bibinfo {author} {\bibfnamefont {Q.}~\bibnamefont {Zhang}},
  \bibinfo {author} {\bibfnamefont {L.}~\bibnamefont {Gu}}, \bibinfo {author}
  {\bibfnamefont {J.}~\bibnamefont {Deng}}, \bibinfo {author} {\bibfnamefont
  {J.}~\bibnamefont {Chen}}, \ and\ \bibinfo {author} {\bibfnamefont
  {X.}~\bibnamefont {Xing}},\ }\href {\doibase
  https://doi.org/10.1002/adma.202002968} {\bibfield  {journal} {\bibinfo
  {journal} {Advanced Materials}\ }\textbf {\bibinfo {volume} {32}},\ \bibinfo
  {pages} {2002968} (\bibinfo {year} {2020})}\BibitemShut {NoStop}%
\bibitem [{\citenamefont {Engelkemier}\ and\ \citenamefont
  {Fredrickson}(2016)}]{PhononsChemistryPV}%
  \BibitemOpen
  \bibfield  {author} {\bibinfo {author} {\bibfnamefont {J.}~\bibnamefont
  {Engelkemier}}\ and\ \bibinfo {author} {\bibfnamefont {D.~C.}\ \bibnamefont
  {Fredrickson}},\ }\href {\doibase 10.1021/acs.chemmater.6b00917} {\bibfield
  {journal} {\bibinfo  {journal} {Chemistry of Materials}\ }\textbf {\bibinfo
  {volume} {28}},\ \bibinfo {pages} {3171} (\bibinfo {year}
  {2016})}\BibitemShut {NoStop}%
\bibitem [{\citenamefont {Osman}\ \emph
  {et~al.}(2018{\natexlab{a}})\citenamefont {Osman}, \citenamefont
  {Salvad{\'o}}, \citenamefont {Pertierra}, \citenamefont {Engelkemier},
  \citenamefont {Fredrickson},\ and\ \citenamefont {Recio}}]{2018ChemicalMaps}%
  \BibitemOpen
  \bibfield  {author} {\bibinfo {author} {\bibfnamefont {H.~H.}\ \bibnamefont
  {Osman}}, \bibinfo {author} {\bibfnamefont {M.~A.}\ \bibnamefont
  {Salvad{\'o}}}, \bibinfo {author} {\bibfnamefont {P.}~\bibnamefont
  {Pertierra}}, \bibinfo {author} {\bibfnamefont {J.}~\bibnamefont
  {Engelkemier}}, \bibinfo {author} {\bibfnamefont {D.~C.}\ \bibnamefont
  {Fredrickson}}, \ and\ \bibinfo {author} {\bibfnamefont {J.~M.}\ \bibnamefont
  {Recio}},\ }\href {\doibase 10.1021/acs.jctc.7b00943} {\bibfield  {journal}
  {\bibinfo  {journal} {Journal of Chemical Theory and Computation}\ }\textbf
  {\bibinfo {volume} {14}},\ \bibinfo {pages} {104} (\bibinfo {year}
  {2018}{\natexlab{a}})}\BibitemShut {NoStop}%
\bibitem [{\citenamefont {Osman}\ \emph
  {et~al.}(2018{\natexlab{b}})\citenamefont {Osman}, \citenamefont
  {Salvad{\'o}}, \citenamefont {Pertierra}, \citenamefont {Engelkemier},
  \citenamefont {Fredrickson},\ and\ \citenamefont {Recio}}]{Osman2018CPmaps}%
  \BibitemOpen
  \bibfield  {author} {\bibinfo {author} {\bibfnamefont {H.~H.}\ \bibnamefont
  {Osman}}, \bibinfo {author} {\bibfnamefont {M.~A.}\ \bibnamefont
  {Salvad{\'o}}}, \bibinfo {author} {\bibfnamefont {P.}~\bibnamefont
  {Pertierra}}, \bibinfo {author} {\bibfnamefont {J.}~\bibnamefont
  {Engelkemier}}, \bibinfo {author} {\bibfnamefont {D.~C.}\ \bibnamefont
  {Fredrickson}}, \ and\ \bibinfo {author} {\bibfnamefont {J.~M.}\ \bibnamefont
  {Recio}},\ }\href {\doibase 10.1021/acs.jctc.7b00943} {\bibfield  {journal}
  {\bibinfo  {journal} {Journal of Chemical Theory and Computation}\ }\textbf
  {\bibinfo {volume} {14}},\ \bibinfo {pages} {104} (\bibinfo {year}
  {2018}{\natexlab{b}})}\BibitemShut {NoStop}%
\bibitem [{\citenamefont {Mellaerts}\ \emph {et~al.}(2022)\citenamefont
  {Mellaerts}, \citenamefont {Seo}, \citenamefont {Afanas'ev}, \citenamefont
  {Houssa},\ and\ \citenamefont {Locquet}}]{BTO2022Simon}%
  \BibitemOpen
  \bibfield  {author} {\bibinfo {author} {\bibfnamefont {S.}~\bibnamefont
  {Mellaerts}}, \bibinfo {author} {\bibfnamefont {J.~W.}\ \bibnamefont {Seo}},
  \bibinfo {author} {\bibfnamefont {V.}~\bibnamefont {Afanas'ev}}, \bibinfo
  {author} {\bibfnamefont {M.}~\bibnamefont {Houssa}}, \ and\ \bibinfo {author}
  {\bibfnamefont {J.-P.}\ \bibnamefont {Locquet}},\ }\href {\doibase
  10.1103/PhysRevMaterials.6.064410} {\bibfield  {journal} {\bibinfo  {journal}
  {Phys. Rev. Mater.}\ }\textbf {\bibinfo {volume} {6}},\ \bibinfo {pages}
  {064410} (\bibinfo {year} {2022})}\BibitemShut {NoStop}%
\bibitem [{\citenamefont {Chiu}\ \emph {et~al.}(2022)\citenamefont {Chiu},
  \citenamefont {Ho}, \citenamefont {Lee}, \citenamefont {Shao}, \citenamefont
  {Shen}, \citenamefont {Liu}, \citenamefont {Chang}, \citenamefont {Zheng},
  \citenamefont {Huang}, \citenamefont {Chang}, \citenamefont {Kuo},
  \citenamefont {Duan}, \citenamefont {Huang}, \citenamefont {Yang},\ and\
  \citenamefont {Chuang}}]{2022UltrathinSTO}%
  \BibitemOpen
  \bibfield  {author} {\bibinfo {author} {\bibfnamefont {C.-C.}\ \bibnamefont
  {Chiu}}, \bibinfo {author} {\bibfnamefont {S.-Z.}\ \bibnamefont {Ho}},
  \bibinfo {author} {\bibfnamefont {J.-M.}\ \bibnamefont {Lee}}, \bibinfo
  {author} {\bibfnamefont {Y.-C.}\ \bibnamefont {Shao}}, \bibinfo {author}
  {\bibfnamefont {Y.}~\bibnamefont {Shen}}, \bibinfo {author} {\bibfnamefont
  {Y.-C.}\ \bibnamefont {Liu}}, \bibinfo {author} {\bibfnamefont {Y.-W.}\
  \bibnamefont {Chang}}, \bibinfo {author} {\bibfnamefont {Y.-Z.}\ \bibnamefont
  {Zheng}}, \bibinfo {author} {\bibfnamefont {R.}~\bibnamefont {Huang}},
  \bibinfo {author} {\bibfnamefont {C.-F.}\ \bibnamefont {Chang}}, \bibinfo
  {author} {\bibfnamefont {C.-Y.}\ \bibnamefont {Kuo}}, \bibinfo {author}
  {\bibfnamefont {C.-G.}\ \bibnamefont {Duan}}, \bibinfo {author}
  {\bibfnamefont {S.-W.}\ \bibnamefont {Huang}}, \bibinfo {author}
  {\bibfnamefont {J.-C.}\ \bibnamefont {Yang}}, \ and\ \bibinfo {author}
  {\bibfnamefont {Y.-D.}\ \bibnamefont {Chuang}},\ }\href {\doibase
  10.1021/acs.nanolett.1c04434} {\bibfield  {journal} {\bibinfo  {journal}
  {Nano Letters}\ }\textbf {\bibinfo {volume} {22}},\ \bibinfo {pages} {1580}
  (\bibinfo {year} {2022})}\BibitemShut {NoStop}%
\bibitem [{\citenamefont {Thomann}(1987)}]{Thomann1987}%
  \BibitemOpen
  \bibfield  {author} {\bibinfo {author} {\bibfnamefont {H.}~\bibnamefont
  {Thomann}},\ }\href {\doibase 10.1080/00150198708227917} {\bibfield
  {journal} {\bibinfo  {journal} {Ferroelectrics}\ }\textbf {\bibinfo {volume}
  {73}},\ \bibinfo {pages} {183} (\bibinfo {year} {1987})}\BibitemShut
  {NoStop}%
\bibitem [{\citenamefont {Cohen}(1992)}]{Cohen1992}%
  \BibitemOpen
  \bibfield  {author} {\bibinfo {author} {\bibfnamefont {R.~E.}\ \bibnamefont
  {Cohen}},\ }\href {\doibase 10.1038/358136a0} {\bibfield  {journal} {\bibinfo
   {journal} {Nature}\ }\textbf {\bibinfo {volume} {358}},\ \bibinfo {pages}
  {136} (\bibinfo {year} {1992})}\BibitemShut {NoStop}%
\bibitem [{\citenamefont {Garcia-Fernandez}\ \emph {et~al.}(2010)\citenamefont
  {Garcia-Fernandez}, \citenamefont {Aramburu}, \citenamefont {Barriuso},\ and\
  \citenamefont {Moreno}}]{CovalencyOctaTilts}%
  \BibitemOpen
  \bibfield  {author} {\bibinfo {author} {\bibfnamefont {P.}~\bibnamefont
  {Garcia-Fernandez}}, \bibinfo {author} {\bibfnamefont {J.~A.}\ \bibnamefont
  {Aramburu}}, \bibinfo {author} {\bibfnamefont {M.~T.}\ \bibnamefont
  {Barriuso}}, \ and\ \bibinfo {author} {\bibfnamefont {M.}~\bibnamefont
  {Moreno}},\ }\href {\doibase 10.1021/jz900399m} {\bibfield  {journal}
  {\bibinfo  {journal} {The Journal of Physical Chemistry Letters}\ }\textbf
  {\bibinfo {volume} {1}},\ \bibinfo {pages} {647} (\bibinfo {year}
  {2010})}\BibitemShut {NoStop}%
\end{thebibliography}
\end{document}